\documentclass[prb,twocolumn,10pt,aps,floatfix,letterpaper,longbibliography,nofootinbib]{revtex4-2}

\usepackage{hyperref}
\usepackage{stix} 
\usepackage{graphicx}
\usepackage{mathtools}

\usepackage{slashed}

\usepackage{braket}
\usepackage{upgreek}
\usepackage{bm}

\usepackage[OT2,T1]{fontenc}
\newcommand{\Sha}{\text{\usefont{OT2}{wncyr}{m}{n}\symbol{"58}}}

\newcommand{\punc}[1]{\,{\text{#1}}}
\newcommand{\sub}[1]{_{\text{#1}}}
\newcommand{\spr}[1]{^{(\text{#1})}}
\newcommand{\super}[1]{^{(#1)}}

\renewcommand{\refeq}[1]{Eq.~(\ref{#1})}
\newcommand{\refeqand}[2]{Eqs.~(\ref{#1}) and (\ref{#2})}
\newcommand{\reffig}[1]{Fig.~\ref{#1}}
\newcommand{\refcite}[1]{Ref.~\cite{#1}}
\newcommand{\refsec}[1]{Section~\ref{#1}}
\newcommand{\refapp}[1]{Appendix~\ref{#1}}

\newcommand{\ee}{\mathrm{e}}
\newcommand{\ii}{\mathrm{i}}
\newcommand{\dd}{\mathrm{d}}
\newcommand{\DD}{\mathrm{D}}

\newcommand{\h}{h}
\newcommand{\J}{J}
\newcommand{\M}{M}

\newcommand{\gp}{\chi}

\newcommand{\idm}{\bm{1}}
\newcommand{\Am}{\mathbf{A}}
\newcommand{\Km}{\mathbf{K}}

\newcommand{\psiv}{\bm{\uppsi}}

\newcommand{\del}{\bm{\partial}}
\newcommand{\zerov}{\bm{0}}
\newcommand{\deltav}{\bm{\delta}}
\newcommand{\rv}{\bm{r}}
\newcommand{\kv}{\bm{k}}
\newcommand{\qv}{\bm{q}}
\newcommand{\av}{\bm{a}}
\newcommand{\uv}{\bm{u}}
\newcommand{\vv}{\bm{v}}
\newcommand{\ev}{\bm{e}}
\newcommand{\nv}{\bm{n}}

\newcommand{\Qv}{\bm{Q}}

\newcommand{\Lv}{\bm{L}}
\newcommand{\Rv}{\bm{R}}
\newcommand{\Bv}{\bm{B}}

\newcommand{\rmM}{\mathrm{M}}

\newcommand{\mv}{\bm{m}}
\newcommand{\phiv}{\bm{\phi}}
\newcommand{\tv}{\bm{t}}
\newcommand{\Phiv}{\bm{\Phi}}
\newcommand{\wv}{\bm{w}}
\newcommand{\pv}{\bm{p}}

\newcommand{\Chi}{\mathrm{X}}

\newcommand{\Dv}{\bm{D}}
\newcommand{\slD}{\slashed{D}}

\newcommand{\tr}{^\intercal}

\newcommand{\scL}{\mathcal{L}}
\newcommand{\scS}{\mathcal{S}}

\newcommand{\goB}{\mathfrak{B}}
\newcommand{\dsZ}{\mathbb{Z}}

\newcommand{\dsC}{\mathbb{C}}

\newcommand{\goz}{\mathfrak{z}}
\newcommand{\Auc}{\mathcal{A}}
\newcommand{\Nuc}{\mathcal{N}}

\newcommand{\detLambda}{\det\nolimits_\Lambda} 
\newcommand{\detLambdainfty}{\det\nolimits_{\Lambda\rightarrow\infty}}

\DeclareMathOperator{\sgn}{sgn}
\DeclareMathOperator{\Tr}{Tr}
\DeclareMathOperator{\trace}{tr}
\DeclareMathOperator{\Pf}{Pf}

\DeclareMathOperator{\re}{Re}
\DeclareMathOperator{\im}{Im}

\usepackage{color}

\makeatletter
\def\beq{\@ifstar{\@ifnextchar[{\@beqslabel}{\@beqsnolabel}}%
{\@ifnextchar[{\@beqlabel}{\@beqnolabel}}}
\def\@beqlabel[#1]{\begin{equation}\label{#1}}
\def\@beqnolabel{\begin{equation}}
\def\@beqslabel[#1]{\begin{equation*}\label{#1}}
\def\@beqsnolabel{\begin{equation*}}
\def\eeq{\@ifstar{\end{equation*}}{\end{equation}}}
\makeatother

\begin{document}

\title{Derivation of height field theory for the two-dimensional classical dimer model from a Grassmann-integral representation}

\author{Stephen Powell}
\affiliation{School of Physics and Astronomy, The University of Nottingham, Nottingham, NG7 2RD, United Kingdom}

\begin{abstract}
The classical dimer model on bipartite lattices hosts a Coulomb phase, characterized by algebraic correlations and topological order. Its long-wavelength properties can be described by the fluctuations of a vector field with zero divergence, which, in two dimensions, is equivalent to a continuum height model. We show how this field theory can be derived constructively for both square and honeycomb lattices, starting from an exact representation of the dimer model in terms of Grassmann integrals. Taking the continuum limit gives a massless Dirac fermion in two-dimensional Euclidean space, which, using bosonization in the path integral representation, maps exactly to the well-known height field theory, incorporating the relationship between its boundary discontinuity (or ``tilt'') and the flux. By including source terms coupling to the flux density and the local valence-bond-solid order parameter, which we show are the only ones required to describe the asymptotic long-distance correlations, we derive expressions for the dimer observables in terms of the height.
\end{abstract}

\maketitle

\section{Introduction}

All physical models are simplifications of reality, but some take this to an extreme, abstracting away as many details as possible to provide insights into the underlying principles. A canonical example is the classical Ising model, which is an oversimplification of real magnetic materials, but has been the setting for the development of much of the modern understanding of phase transitions and critical phenomena, and finds applications in many other areas of physics and beyond \cite{Brush1967,Fisher1981,Kulske2025}.

In the same spirit, dimer models \cite{KenyonLesHouches} can be viewed as maximally simplified systems in which to study the unconventional collective behavior that can occur in many-body systems with strong local constraints. In these models, a dimer is an object that occupies a pair of adjacent sites of a lattice and, if one restricts to close-packed configurations, each site is occupied by exactly one dimer. Dimer models were originally introduced to describe the arrangement of diatomic molecules on the surface of a crystal \cite{Roberts1935,Fowler1937,Roberts1939}, and have also been applied as descriptions of frustrated magnets \cite{Moessner2003,Shannon2010,Chalker2017}, liquid crystals \cite{Heilmann1979,Jauslin2018}, molecular tilings \cite{Blunt2008}, and mesoscale structures \cite{Meng2025}. From a more theoretical point of view, they provide an ideal venue to study the phenomena that local constraints can give rise to, such as effective gauge theories \cite{Huse2003}, topological order \cite{Henley2010,Henley2011}, non-Landau phase transitions \cite{Kasteleyn1963,Alet2005,Alet2006,Sreejith2019}, and cooperative relaxation \cite{Henley1997,Oakes2016,Feldmeier2021}.

On many bipartite lattices, including square and honeycomb, an equal-weight ensemble of all close-packed dimer configurations is a Coulomb phase \cite{Henley2010}, in which correlations decrease algebraically at large separation. The long-distance physics can be captured by rewriting the close-packing constraint as a lattice Gauss law for a suitably defined ``magnetic flux density'' \cite{Huse2003} and conjecturing a continuum theory in terms of a corresponding coarse-grained vector field. One can then resolve the Gauss law in two dimensions (2D) using a height field \cite{Blote1982,Nienhuis1984}, giving a sine-Gordon model (or, equivalently, a Coulomb gas) \cite{Nienhuis1987}, and in 3D using a vector potential, giving a noncompact \(\mathrm{U}(1)\) gauge theory \cite{Huse2003}. These field theories are well established as continuum descriptions of both 2D and 3D classical dimer models, including their dynamics \cite{Henley1997,Moessner2003}, and have also been used as starting points for understanding the quantum dimer model \cite{Rokhsar1988,Moessner2011} at and near the Rokhsar--Kivelson point \cite{Fradkin2004,Patil2014,Ardonne2004}.

A different approach is possible in 2D, based on the fact that the partition function of the dimer model can be calculated exactly on any planar graph. This can be done either by expressing it in terms of the Pfaffian of an appropriately constructed matrix \cite{Kasteleyn1961,Fisher1961,Temperley1961} or using a transfer matrix and Jordan--Wigner transformation to express it in terms of free fermions \cite{Lieb1967}. As well as planar graphs, which correspond to lattices with closed (or cylindrical) boundary conditions, the solution can be extended to periodic boundary conditions, as well as to surfaces of higher genus \cite{Cimasoni2007}. Dimer correlations can be calculated using the exact solutions \cite{Fisher1963} and have been used to fix the coefficients in the continuum height theory \cite{Blote1982,Henley1997}.

In \refcite{Wilkins2023}, it was shown how the two approaches could be brought together to derive the continuum height theory constructively, starting from the transfer-matrix solution on the square lattice. Expanding around the Fermi points of the fermionic dispersion \cite{Wilkins2021} to arrive at a continuum description and then using bosonization \cite{vonDelft1998}, one finds a continuum model in terms of bosonic operators. This can be expressed as a real field theory in space and imaginary time by the standard path-integral mapping, thereby restoring the isotropic nature of the original model. The result agrees with the standard height theory, including the correspondences between microscopic dimer observables and operators in the field theory that were posited previously.

Here, we show how the same theory can be derived using a more direct method that treats both dimensions on an equal footing throughout. We start with a Grassmann-integral representation of the partition function \cite{Samuel1980}, which is then  expanded in terms of slowly varying fermion modes. The resulting continuum description is exactly a Dirac fermion in 2D Euclidean space, which maps to the required real field theory using bosonization identities in the field-theoretical formulation. We use periodic boundary conditions, which imply a sum over boundary phases (or ``spin sectors'') for the fermions, and show how these lead to tilted boundary conditions for the continuum height. In addition, we include source terms that couple to the local flux density and to the  valence-bond-solid (VBS) order parameter, a different local combination of the dimer degrees of freedom. These map to a \(\mathrm{U}(1)\) gauge field and a mass source, respectively, for the Dirac fermion, and to the gradient and complex exponential (``vertex operator'') of the bosonic field. We furthermore demonstrate that these sources are the only ones required to describe the long-wavelength modes of the model.

Compared to the approach based on the transfer matrix, this method has the advantage of keeping real space explicit, and retaining the full lattice structure until the continuum limit is taken. Here, we consider both square and honeycomb lattices and show that an identical continuum theory is found for both.

In \refsec{SecDimerModel}, we define the classical dimer model and the partition function including sources. We then express this in terms of Grassmann integrals in \refsec{SecGrassmannIntegrals} and review the calculation of the partition function with the sources set to zero. We derive a continuum theory in terms of Dirac fermions in \refsec{SecDiracFermions} and then use bosonization to map this to a bosonic field theory in \refsec{SecBosonization}.

\section{Dimer model}
\label{SecDimerModel}

We consider a classical statistical model involving dimers on the links \(\ell\) of a lattice, with occupation number \(d_\ell = 0\) or \(1\), and restrict to close-packed configurations, without holes (monomers) or overlapping dimers. In other words, exactly one dimer touches each site \(i\),
\beq[EqClosePacking]
\sum_{\ell \in i} d_\ell = 1
\eeq
where the sum is over all links \(\ell\) connected to \(i\). We are primarily interested in correlation functions of \(d_\ell\) between different links, in the statistical ensemble where of all such configurations have equal weight.

In this section, we define the flux density \(B_\ell\) and VBS order parameter (or ``magnetization'') \(\Psi_i\), which are local functions of the dimer variables \(d_\ell\). We then define a partition function for the dimer model including sources that couple to these quantities, from which arbitrary correlation functions can be calculated.

\subsection{Lattice structure}
\label{SecLatticeStructure}

We restrict to 2D, where this problem can be solved exactly, and to bipartite lattices, where a height mapping can be applied. For simplicity, we furthermore assume that one can define a unit cell containing one site of each sublattice, although similar methods can be applied to more general lattices. These conditions in fact restrict the analysis to the square and honeycomb lattices, but we use general notation to allow both to be treated.

\begin{figure}
\includegraphics{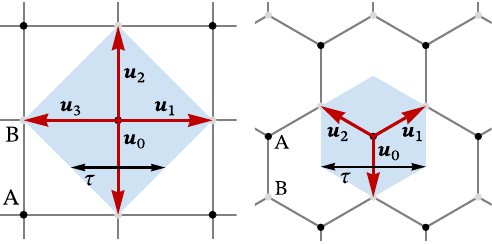}
\caption{Nearest-neighbor vectors \(\uv_b\) for the square (left) and honeycomb (right) lattices. In each case, the two sublattices, A and B, are labeled and the shaded polygon shows a unit cell containing one site of each sublattice. (For the square lattice, this is twice the area of the usual unit cell.) The black horizonal arrows show the distance \(\tau\) between adjacent plaquette centers, given by \(\tau = \lvert \uv_b \rvert\) for the square lattice and \(\tau = \sqrt{3}\lvert \uv_b \rvert\) for honeycomb.}
\label{FigVectors}
\end{figure}
A bipartite lattice is one whose sites can be divided into sublattices \(\mathrm{A}\) and \(\mathrm{B}\) such that each site \(i \in \mathrm{A}\) has all neighbors \(j \in \mathrm{B}\), and vice versa. We denote by \(\ell = ib \equiv jb\) the link between neighboring sites \(i\) and \(j\) at positions \(\rv_i\) and \(\rv_j = \rv_i + \uv_b\), with \(b = 0,1, \dotsc,\goz-1\), where \(\goz\) is the coordination number. The nearest-neighbor vectors \(\uv_b\) have equal length \(u = \lvert\uv_b\rvert\) and are separated by equal angles \(2\pi/\goz\).\footnote{The dimer model can be defined on a graph, but the mapping to a continuum theory relies on an embedding in the plane. For example, the dimer model is equivalent on the honeycomb and brick lattices \cite{Yokoi1986}, but these definitions assume the embedding for the honeycomb lattice.} As shown in \reffig{FigVectors}, we number them counterclockwise starting from \(\uv_0 = -u\deltav_y\), where \(\deltav_\mu\) (for \(\mu\in\{x,y\}\)) is a Cartesian unit vector. They form an overcomplete basis for 2D vectors, obeying the completeness relation
\beq[EquuSum]
\sum_b (\uv_{b})_\mu (\uv_b)_\nu = \frac{1}{2}\goz u^2 \delta_{\mu\nu}
\punc.
\eeq

We use periodic boundary conditions, which, for simplicity, are applied along the Cartesian axes, with lengths \(L_x\) and \(L_y\). Defining vectors \(\Lv_\mu = L_\mu \deltav_\mu\) (no summation over \(\mu\)), the positions \(\rv\) and \(\rv + \Lv_\mu\) are equivalent. For the square lattice to remain bipartite, \(L_x\) and \(L_y\) must both be even multiples of the nearest-neighbor distance \(u\). For the honeycomb lattice, \(L_x\) and \(L_y\) should be integer multiples of \(\sqrt{3}u\) and \(3u\) respectively. (We apply an additional condition on \(L_x\) in \refsec{SecMagnetization}.)

\subsection{Flux and height}
\label{SecDiscreteHeight}

\begin{figure*}
\includegraphics{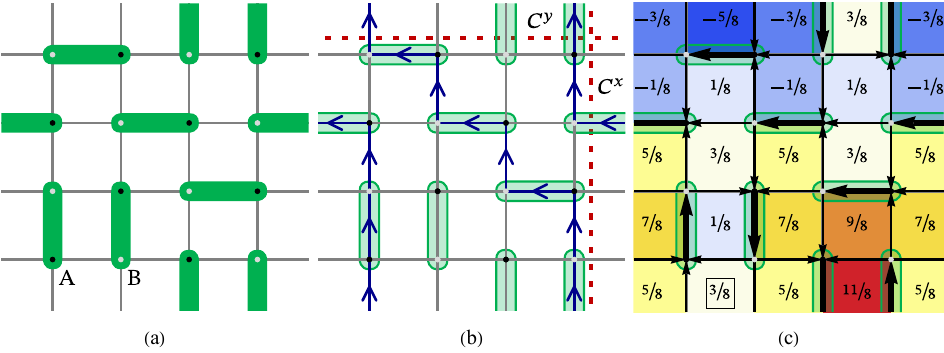}
\caption{(a) A configuration \(c\) of the dimer model on a \(4\times 4\) square lattice with periodic boundary conditions. The two sublattices, A and B, of the square lattice are labeled. (b) Corresponding directed-loop configuration (see \refsec{SecDirectedLoops}), shown using blue arrowheads, constructed using the reference configuration \(c_0\) and directed Kasteleyn graph of \reffig{FigStaggeredAndKasteleyn}. Edges appear on links for which the dimer occupation in \(c\) and \(c_0\) is not equal, with direction as in the graph. The dashed red lines show surfaces \(C^x\) and \(C^y\) that span the periodic boundaries (i.e., loops that wind around the handles of the torus), with normals taken in the positive Cartesian directions. (c) Configuration of flux density \(B_\ell\) (arrows) and height \(h_p\) for \(c\). Large and small arrows represent \(B_\ell = \frac{3}{4}\) and \(\frac{1}{4}\) respectively, and so the divergence of \(B_\ell\) (outgoing minus incoming flux) is zero on every site. The height is defined by \refeq{EqDefineHeight}; the arrows act as its (right-handed) contours, surrounding peaks in a counterclockwise sense. The overall level of the height is arbitrary and has been fixed to \(h_0 = \frac{3}{8}\) on a plaquette (shown with a rectangle) near the bottom-left corner. With this choice, the magnetization (see \refsec{SecMagnetization}) is given by \(\Psi_i = \ee^{-2\pi \ii h_i}\) in any configuration, where \(h_i\) is the mean of \(h_p\) on plaquettes surrounding site \(i\). This configuration has flux \(\Phiv = (-1,0)\), and so the height is periodic in the horizontal direction but has a step around the periodic boundaries in the vertical direction, \(h(\rv + \Lv_y) = h(\rv) - 1\); see \refeq{EqFluxTilt}.}
\label{FigRandomLoopsHeight}
\end{figure*}

\begin{figure*}
\includegraphics{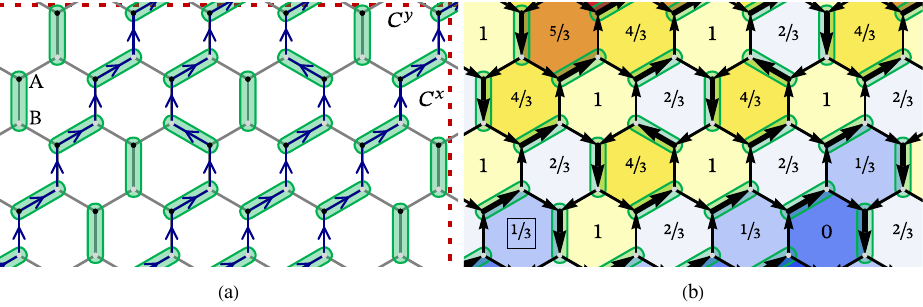}
\caption{(a) A configuration of the dimer model on the honeycomb lattice, with the corresponding directed-loop configuration (see \refsec{SecDirectedLoops}) shown using blue arrowheads. The dashed red lines show the surfaces \(C^x\) and \(C^y\), as in \reffig{FigRandomLoopsHeight}(b). (b) Configuration of flux density \(B_\ell\) (arrows) and height \(h_p\) for \(c\). Large and small arrows represent \(B_\ell = \frac{2}{3}\) and \(\frac{1}{3}\) respectively. The height has been fixed to \(h_0 = \frac{1}{3}\) on a plaquette near the bottom-left corner.}
\label{FigRandomLoopsHeightHc}
\end{figure*}

The topological structure of the dimer ensemble and the mapping to the microscopic height variable can both be understood starting from the flux density (or effective magnetic field). This is defined for any dimer configuration by
\beq[EqDefineB]
B_{\ell} = d_{\ell} - \frac{1}{\goz}\punc,
\eeq
which we treat as directed out of the sublattice-A sites and into sublattice-B sites. The constant is added so that its lattice divergence \cite{GradyPolimeni2010},
\beq[EqDefineDiv]
\operatorname{div}_i B = \eta_i \sum_{\ell \in i} B_\ell\punc,
\eeq
where \(\eta_i = +1\) for \(i \in \mathrm{A}\) and \(-1\) for \(i\in\mathrm{B}\), obeys the Gauss law \(\operatorname{div}_i B = 0\) for all \(i\), in any dimer configuration obeying the close-packing constraint, \refeq{EqClosePacking}. The flux density is illustrated for typical dimer configurations by the arrows in \reffig{FigRandomLoopsHeight}(c) and \ref{FigRandomLoopsHeightHc}(b).

Let \(C^x\) and \(C^y\) be surfaces (i.e., loops) that span the periodic boundaries in the \(y\) and \(x\) directions respectively, e.g., the dashed red lines in \reffig{FigRandomLoopsHeight}(b) and \ref{FigRandomLoopsHeightHc}(a). The flux \(\Phiv\) is a vector with components
\beq[EqDefinePhi]
\Phi_\mu = \sum_{\ell} \dsC^\mu_\ell B_{\ell}\punc,
\eeq
where \(\dsC^\mu_\ell = \pm 1\) if \(\ell\) (directed from A to B) crosses \(C^\mu\) in a positive or negative direction, and \(\dsC^\mu_\ell = 0\) otherwise.  Because \(B_\ell\) has zero divergence, \(\Phiv\) is invariant both under local dimer rearrangements and under local deformations of the surfaces \cite{Chalker2017}. Using this topological invariance,\footnote{Take \(C^\mu\) as a straight line along Cartesian direction \(\mu\) and average over all possible positions in the transverse direction (excluding the measure-zero set where \(C^\mu\) passes through a lattice site). This gives \(\overline{\dsC^\mu_\ell} = (\uv_{b_\ell})_\mu /L_\mu\), where \((\uv_{b_\ell})_\mu = \uv_{b_\ell} \cdot \deltav_\mu\) is the projection along \(\mu\) of the nearest-neighbor vector \(\uv_{b_\ell}\) for link \(\ell\).} one can derive an alternative,  global expression for the flux,
\beq[EqDefinePhi2]
\Phi_\mu = \frac{1}{L_\mu}\sum_{\ell} B_{\ell} \big(\uv_{b_\ell}\big)_\mu
\punc.
\eeq

The zero-divergence constraint on \(B_\ell\) can be resolved by defining the ``height'' \(h_p\) \cite{Nienhuis1984}. This is assigned to each plaquette \(p\) of the lattice such that
\beq[EqDefineHeight]
B_\ell = \Delta_{\perp \ell} h \equiv h_{p\sub{L}} - h_{p\sub{R}}
\punc,
\eeq
where \(\Delta_{\perp\ell}\) is the difference of the values on the plaquettes to the left \(p\sub{L}\) and right \(p\sub{R}\) of the (directed) link \(\ell\). This is the discrete analogue of the 2D curl \(\epsilon_{\mu\nu}\partial_\nu\) of a continuum scalar \cite{GradyPolimeni2010}, where \(\epsilon_{\mu\nu}\) is the Levi-Civita tensor and summation over the repeated index is implied (here and in subsequent expressions involving \(\epsilon_{\mu\nu}\)). This definition determines \(h_p\) only up to a global shift, which is resolved by fixing its value at an arbitrarily chosen plaquette \(p_0\). The height is shown for example configurations in \reffig{FigRandomLoopsHeight}(c) and \ref{FigRandomLoopsHeightHc}(b). By construction, the \(B_\ell\) arrows are contours of the height, surrounding peaks in a counterclockwise direction.

Combining \refeqand{EqDefinePhi}{EqDefineHeight} and summing successive height differences across the surface \(C^\mu\), one finds
\beq[EqFluxTilt]
\Phi_\mu = \epsilon_{\mu\nu}\left[ h(\rv + \Lv_\nu) - h(\rv) \right]\punc,
\eeq
where \(h(\rv)\) is the height on a plaquette centered at \(\rv\). The height therefore has a discontinuity around the boundary conditions determined by the transverse component of the flux.\footnote{\label{FootnoteWindingNumber}The winding number or ``tilt'' is often defined as \(W_\mu = h(\rv + \Lv_\mu) - h(\rv)\) \cite{Henley1997} and is related to the flux by \(\Phi_\mu = \epsilon_{\mu\nu}W_\nu\).}

\subsection{Magnetization}
\label{SecMagnetization}

\begin{figure*}
\centering
\includegraphics{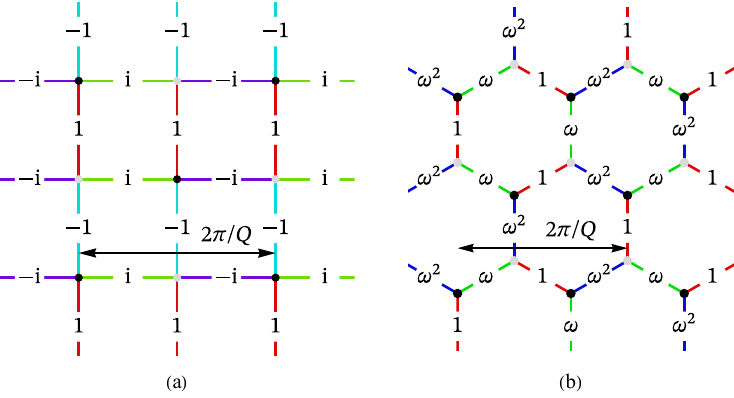}
\caption{Link phases \(\vartheta_\ell\) used to define the magnetization (VBS order parameter) for (a) the square lattice and (b) the honeycomb lattice, with \(\omega = \ee^{2\pi \ii /3}\). For both, the sublattice-A site at the bottom left is the origin. This pattern can be associated with a wavevector \(\Qv = (Q, 0)\); the arrows show the corresponding spatial periods \(2\pi/Q\).}
\label{FigVbsPhases}
\end{figure*}

To characterize long-range correlations in the dimer model, we also use the complex VBS order parameter \(\Psi_i\) \cite{Sachdev1989,Pujari2015}, which we will refer to as the magnetization. This quantity is constructed such that \(\left\lvert \sum_i \Psi_i\right\rvert\) is large in configurations that maximize the number of flippable plaquettes (so-called ``ideal states'' \cite{Kondev1996,Raghavan1997}). It is therefore used as an order parameter for certain ordering transitions \cite{Alet2006b,Alet2006,Sreejith2019}, including in quantum dimer models \cite{Sachdev1989,Patil2014,Oakes2018,Yan2021}.

We define it as a complex scalar on each site \(i\),\footnote{On the square lattice, the magnetization can instead be defined as a real quantity on directed links \cite{Wilkins2020}, as with \(B_\ell\). For honeycomb, however, the corresponding ordered states are not collinear, and so a complex number is more convenient.}
\beq[EqDefinePsi]
\Psi_i = \sum_{\ell \in i} d_\ell \vartheta_\ell\punc,
\eeq
where \(\vartheta_\ell\) is a fixed complex phase (i.e., \(\lvert\vartheta_\ell\rvert = 1\)) defined on the links, illustrated in \reffig{FigVbsPhases}. These phases are assigned such that \(\vartheta_\ell\) winds uniformly once around the unit circle when going around an A-sublattice site in the same sense, or a B-sublattice site in the opposite sense. We fix the overall phase by setting \(\vartheta_\ell=1\) on the link (with \(b=0\)) pointing vertically downwards from the site at the origin.

One can show that consistency between next-neighbor A-sublattice sites separated by unit vectors \(\ev_b = \uv_b - \uv_0\) requires that
\beq[EqVartheta]
\vartheta_{ib} = \ee^{-\ii \Qv \cdot \rv_i} \omega^{\eta_i b}\punc,
\eeq
where \(\omega = \ee^{2\pi \ii/\goz}\) and \(\Qv\) obeys
\beq[EqDefineQ]
\ee^{-\ii \Qv \cdot (\uv_b - \uv_0)} = \omega^{2b}\punc,
\eeq
for all \(b\). For both the square and honeycomb lattices, this can be satisfied by choosing \(\Qv = (Q,0)\) along \(x\), with \(Q = 2\pi u / \Auc\), where \(\Auc = \ev_1 \times \ev_2\) is the area of the unit cell (containing one site of each sublattice).

The corresponding spatial periods, \(2\pi/Q=2u\) on the square lattice and \(\frac{3\sqrt{3}}{2}u\) on the honeycomb lattice, are also shown in \reffig{FigVbsPhases}. To apply the phases consistently throughout the lattice, we require \(L_x\) to be an integer multiple of \(2\pi/Q\). For honeycomb, we have already required in \refsec{SecLatticeStructure} that \(L_x\) be a multiple of \(\sqrt{3}u\); together these imply that it is a multiple of \(3\sqrt{3}u\), as is clear in \reffig{FigVbsPhases}(b). For the square lattice, it is sufficient for \(L_x\) to be an even multiple of \(u\).

The magnetization \(\Psi_i\) has a simple relationship with the height \(h_p\), as we now show. First note that, according to its definition in \refeq{EqDefineHeight}, \(h_p\) either steps down by \(1/\goz\) (if \(d_\ell = 0\)) or up by \(1 - 1/\goz\) (if \(d_\ell = 1\)) when crossing link \(\ell\) with the A-sublattice site on the left. It follows that \(\ee^{-2\pi \ii h_p}\) winds around the unit circle when going around any site, in the same way as \(\vartheta_\ell\). This fixes \(\ee^{-2\pi \ii h_p}\) equal to \(\vartheta_\ell\) up to a global phase, which we are free to fix and so cannot depend on the configuration. We can therefore write \(\ee^{-2\pi \ii h_{p\sub{L}}}=\ee^{-2\pi \ii h_{p_0}}\vartheta_\ell\), where \(p\sub{L}\) is the plaquette to the left of link \(\ell\) and \(p_0\) is the plaquette immediately down and to the right of the origin (using our convention for the overall phase of \(\vartheta_\ell\)).

Now consider a site \(i\) with its dimer on link \(\ell = ib\). Starting on \(p\sub{L}\) and going counterclockwise around \(i\) if \(i \in \mathrm{A}\) (or clockwise if \(i \in \mathrm{B}\)), the heights step down from \(h_{p\sub{L}}\) to \(h_{p\sub{L}} - (\goz-1)/\goz\). The mean height on the \(\goz\) plaquettes surrounding site \(i\) is therefore \(\h_i = h_{p\sub{L}} - \frac{1}{2}(\goz-1)/\goz\). If we choose \(h_{p_0} = \frac{1}{2}(\goz-1)/\goz\) up to an integer, then \(\ee^{-2\pi \ii \h_i} = \vartheta_\ell\); with this choice, we have
\beq[EqPsiViaHeight]
\Psi_i = \ee^{-2\pi \ii \h_i}
\punc.
\eeq
It is straightforward to check this result for the configurations shown in \reffig{FigRandomLoopsHeight}(c) and \ref{FigRandomLoopsHeightHc}(b), in which the height is fixed according to this convention, by comparing with \reffig{FigVbsPhases}. (The same identity is illustrated for the triangular lattice Ising antiferromagnet, equivalent to the honeycomb dimer model, in \refcite{Zeng1997}.)

Because of the complex phases \(\vartheta_\ell\), the magnetization \(\Psi_i\) transforms under a projective representation of the space group \cite{Wen2002}, and fixing the height so that it satisfies \refeq{EqPsiViaHeight} implies that \(h_i\) also transforms nontrivially \cite{Alet2006b}.\footnote{Specifically, translation by a lattice vector \(\Rv\) shifts \(h_i\) by \(\frac{1}{2\pi}\Qv\cdot \Rv\), rotation by an angle \(2\pi/\goz\) around an A-sublattice site shifts \(h_i\) by \(-1/\goz\), and reflection in a vertical line passing through the origin gives \(h_i \rightarrow -h_i\). On the square lattice, translation by \(\uv_0\) (exchanging sublattices) is also a symmetry and gives \(h_i \rightarrow \frac{1}{2}-h_i\).} Note that fixing \(h_p\) on an arbitrary plaquette \(p_0\) does not imply that the mean height \(h_i\) is fixed on any site. In fact, symmetry under rotation around site \(i\) implies \(\langle\Psi_i\rangle = 0\) in the ensemble of all dimer configurations.

\subsection{Partition function including sources}
\label{SecPartitionFunctionSources}

We can now define the partition function as
\beq[EqDefineZ]
Z = \sum_c \exp \left[
\ii \sum_\ell \alpha_\ell B_\ell + \sum_{i} (\Psi_i^* J_i + \Psi_i J_i^*)
\right]\punc,
\eeq
a sum over all close-packed dimer configurations \(c\) [i.e., those obeying \refeq{EqClosePacking}], weighted by sources \(\alpha_\ell\) and \(J_i\). (The factor of \(\ii\), included with \(\alpha_\ell\) but not \(\J_i\), is arbitrary and chosen to match conventions used elsewhere.) Because \(B_\ell\) has zero lattice divergence, there is a gauge invariance under shifts of \(\alpha_\ell\) by \(\Delta_\ell \Chi\), the lattice gradient of \(\Chi_i\) defined on sites \(i\).

The lattice Helmholtz decomposition \cite{GradyPolimeni2010} allows one to write
\beq[EqLatticeHelmholtz]
\alpha_\ell = a_\ell + \sum_\mu \frac{t_\mu}{L_\mu} \uv_{b_\ell} \cdot \deltav_\mu\punc,
\eeq
where \(a_\ell = \Delta_{\perp\ell} \beta + \Delta_\ell \gp\) is the sum of a lattice curl and gradient, and the second term is a ``uniform'' part.\footnote{Because \(\sum_b \uv_b = \zerov\), this term has zero lattice divergence, defined by \refeq{EqDefineDiv}, and zero lattice curl, defined as the sum over directed links \(\ell\) belonging to a plaquette \(p\) \cite{GradyPolimeni2010}.}  Using \refeq{EqDefinePhi2}, the coupling to \(B_\ell\) is
\beq[EqalphaB]
\sum_\ell \alpha_\ell B_\ell = \sum_\ell a_\ell B_\ell + \tv \cdot \Phiv\punc.
\eeq
It is convenient to apply a gauge transformation, to replace \refeq{EqLatticeHelmholtz} by
\beq[Eqalphaat]
\alpha_\ell = a_\ell + \sum_\mu t_\mu\dsC^\mu_\ell\punc,
\eeq
in which \(t_\mu\) is applied only at the surface \(C^\mu\). Comparison of \refeqand{EqDefinePhi}{EqDefinePhi2} shows that this leaves \refeq{EqalphaB} unchanged. The fact that \(a_\ell\) can be written as \(a_\ell = \Delta_{\perp\ell} \beta + \Delta_\ell \gp\) is equivalent to
\beq[EqaCondition]
\sum_\ell \uv_{b_\ell} a_\ell = \zerov
\punc,
\eeq
which can be treated as the defining condition for the decomposition in \refeq{Eqalphaat}. One can always use the gauge invariance to set \(\gp_i = 0\) on all sites, giving the ``Lorenz gauge``, \(\operatorname{div}_i a = 0\), but we do not assume that this has been done.

We can therefore write the partition function as
\beq[EqZah]
Z(\tv,a,\J) = \sum_c \ee^{\ii \tv \cdot \Phiv}
\ee^{S_c(a,\J)},
\eeq
where
\beq[EqScah]
S_c(a,\J) = \ii \sum_\ell a_\ell B_\ell + \sum_{i} (\Psi_i^* J_i + \Psi_i J_i^*)
\eeq
is the coupling to the sources. The latter will be written as \(S_c\) where there is no ambiguity.

The coupling can be expressed in terms of the dimer occupation using \refeqand{EqDefineB}{EqDefinePsi},
\beq[EqScah2]
S_c = \sum_\ell d_\ell \left(\ii a_\ell + 2\vartheta_\ell^* \J_\ell+ 2\vartheta_\ell \J_\ell^*
\right)
\punc,
\eeq
where \(J_\ell=\frac{1}{2}\sum_{i\in\ell} J_i\) is the mean of \(J_i\) on the two sites connected by the link \(\ell\), and in terms of the height using \refeqand{EqDefineHeight}{EqPsiViaHeight},
\beq[EqScHeight]
S_c = \ii \sum_\ell a_\ell \Delta_{\perp \ell} h + \sum_{i} \left(\ee^{+2\pi \ii \h_i} J_i + \ee^{-2\pi \ii \h_i} J_i^*\right)
\punc.
\eeq

Dimer correlations can be calculated by taking derivatives of \(Z(\tv,a,\J)\) with respect to \(a_\ell\) and \(\J_i\). Our aim is to define a continuum theory that reproduces long-distance correlations \(\braket{d_{\ell_1}d_{\ell_2}\dotsm d_{\ell_n}}\), where the spatial separation between each pair of links is much larger than the lattice spacing. For this purpose, it is sufficient to take \(a\) and \(\J\) as lattice discretizations of a continuum vector field \(\av(\rv)\) and a continuum complex scalar field \(\M(\rv)\), both of which are slowly varying on the lattice scale, writing
\begin{align}
\label{EqDefineav}
a_{ib} &= \uv_b\cdot\av(\rv_i)\\
\J_{i} &= \frac{1}{4}u\M(\rv_i)\punc.
\label{EqDefinescJ}
\end{align}
This choice for the source terms and their relationship to the continuum fields, i.e., the dependence on \(b\) in \refeqand{EqDefineav}{EqDefinescJ}, is justified in \refsec{SecExpansion}, where we show that these are the only sources required to describe the leading-order long-distance correlations. (The coefficients in their definitions are arbitrary, but are chosen to simplify the continuum theory.) Including both \(a_\ell\) and \(\J_i\) in the microscopic partition function is redundant, but the restriction to slowly-varying sources means that both are required.

Note that, using \refeq{EquuSum}, the property \refeq{EqaCondition} implies
\beq[EqaCondition2]
\int \dd^2\rv\, \av(\rv) = \zerov\punc,
\eeq
and so one can write \(a_\mu(\rv) = \epsilon_{\mu \nu}\partial_\nu \beta(\rv) + \partial_\mu \gp(\rv)\), where \(\beta(\rv)\) and \(\gp(\rv)\) are scalars that can be chosen smooth and periodic. The gauge invariance of \(\alpha_\ell\) implies that the partition function is invariant under continuum gauge transformations \(\av(\rv) \rightarrow \av(\rv) + \del \Chi(\rv)\), and so \(\gp(\rv)\) is redundant.

The coupling to slowly varying sources can instead be expressed in terms of coarse-grainings of the microscopic variables \(B_\ell\) and \(\Psi_i\). For sufficiently smooth \(\av\) and \(\M\), \refeq{EqScah} becomes
\beq[EqScahSmoothed]
S_c \approx \int \dd^2 \rv\,
\left[
\ii \av(\rv) \cdot \Bv(\rv)
+ \Psi^*(\rv)\M(\rv) + \Psi(\rv)\M^*(\rv)
\right]
\punc,
\eeq
where
\begin{align}
\label{EqBrv}
\Bv(\rv)&\approx\frac{\goz}{\Auc}\uv_{b_\ell} B_{\ell}\\
\Psi(\rv)&\approx\frac{u}{2\Auc}\Psi_i = \frac{u}{2\Auc}\ee^{-2\pi \ii \h_i}
\label{EqPsirv}
\punc.
\end{align}
In these expressions, ``\(\approx\)'' means that the left-hand side is equal to an average of the right-hand side over sites \(i\) or links \(\ell\) in a neighborhood of \(\rv\) much larger than the lattice scale. We similarly define a coarse-grained height field \(\h(\rv)\approx h_p\), or equivalently \(h(\rv) \approx h_i\); using \(\Auc = \frac{1}{2}\goz u \tau\), where \(\tau\) is the distance between plaquette centers,\footnote{The dual plaquette corresponding to each site is a regular polygon with \(\goz\) sides of length \(\tau\) and apothem \(\frac{1}{2}u\); its area is therefore \(\frac{1}{4}\goz u \tau\). The area of a unit cell containing two sites, is twice this, \(\Auc = \frac{1}{2}\goz u \tau\).} one can show\footnote{The dual link between the centers of plaquettes \(p\sub{L}\) and \(p\sub{R}\) has displacement \((\rv_{p\sub{L}} - \rv_{p\sub{R}})_\mu = -\frac{\tau}{u}\epsilon_{\mu\nu}(\uv_b)_\nu\), where \(b\) is the orientation of the corresponding direct link, and so, for slowly varying \(h(\rv)\), we have \(\uv_b \times \del h(\rv) \approx \frac{u}{\tau}\big[h(\rv_{p\sub{L}}) - h(\rv_{p\sub{R}})\big]\). The completeness relation, \refeq{EquuSum}, can be used to write \(\epsilon_{\mu\nu}\partial_\nu h(\rv) = \frac{2}{\goz u^2} \sum_b (\uv_b)_\mu \uv_b \times \del h(\rv)\). Summing over the \(\goz\) links in a unit cell gives the right-hand side of \refeq{EqBrv}.} that \(B_\mu(\rv) = \epsilon_{\mu\nu}\partial_\nu h(\rv)\). The goal of the following sections, realized in \refeq{EqZahBosonic}, is to derive a distribution for the continuum height \(h(\rv)\) which correctly reproduces all long-range correlation functions for the dimer model.

Finally, using \refeq{EqDefineB} for \(B_\ell\) and \refeqand{EqDefinePsi}{EqVartheta} for \(\Psi_i\), the coarse-grained observables can be expressed in terms of the dimer occupations as
\begin{align}
\label{EqBvfromd}
\Bv(\rv) &\approx \frac{\goz}{\Auc}\uv_{b} \left(d_{ib}-\frac{1}{\goz}\right)\\
\Psi(\rv) &\approx \frac{u \goz}{2\Auc} d_{ib} \ee^{-\ii \Qv\cdot \rv_i}\omega^{\eta_i b}
\label{EqPsifromd}
\punc.
\end{align}
These expressions are used to write the dimers in terms of the continuum fields in \refsec{SecContinuumHeightTheory}.

\section{Dimer partition function in terms of Grassmann integrals}
\label{SecGrassmannIntegrals}

In this section, we rewrite the weighted partition function \(Z(a,\J)\) exactly in terms of an integral over a set of Grassmann variables, following the general method introduced in \refcite{Samuel1980}.

\subsection{Mapping to directed loops}
\label{SecDirectedLoops}

We begin by defining a mapping from dimer configurations to configurations of directed loops. The difference between any two configurations of a dimer model, i.e., the set of links that are occupied in one but not both, forms an arrangement of closed, nonintersecting loops. We perform the mapping by choosing a fixed reference configuration \(c_0\) to which all other configurations \(c\) are compared, which gives a one-to-one mapping between dimer and loop configurations. We assign to each loop a direction, pointing from sublattice A to B on links occupied in \(c\) and from B to A for those occupied in \(c_0\). The choice of a fixed reference configuration implies that only one direction is possible on each link; a given loop configuration either includes this link with this orientation, or not at all.

\begin{figure}
\includegraphics{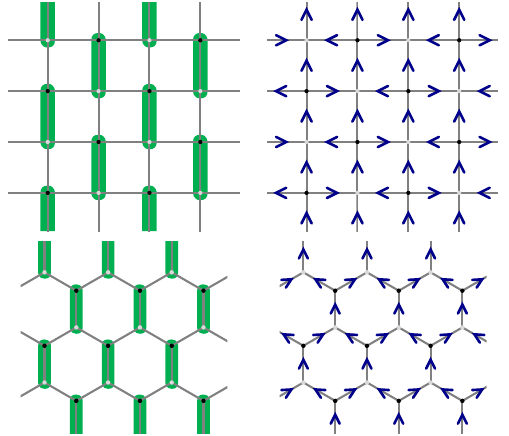}
\caption{Reference configuration \(c_0\) (left column) and corresponding directed Kasteleyn graph (right) on the square lattice (top row) and honeycomb lattice (bottom). Sublattices A and B are shown with black and light-gray points respectively, as labeled in \reffig{FigRandomLoopsHeight}(a) and \ref{FigRandomLoopsHeightHc}(a). Lattice sites are vertices of the graph and each link gives a single directed edge. For links that are unoccupied in the reference configuration, the edge is directed from sublattice A to sublattice B. For links that are occupied, the edge is reversed. The loop configuration for a given dimer configuration \(c\), such as shown \reffig{FigRandomLoopsHeight}(b) and \ref{FigRandomLoopsHeightHc}(a), is given by the subset of the edges where \(c\) and \(c_0\) differ.}
\label{FigStaggeredAndKasteleyn}
\end{figure}

For the square and honeycomb lattices, we choose the reference configurations shown in \reffig{FigStaggeredAndKasteleyn}, with all links with \(b=0\) occupied and all others unoccupied. These have the property that no local rearrangements of dimers are possible, and so any other configuration is reached by shifting a set of dimers that spans the boundaries. This therefore constrains the topology of the loop configurations, which, as we will show, simplifies the mapping to Grassmann integrals.

The corresponding link directions, also shown in \reffig{FigStaggeredAndKasteleyn}, are interpreted as directed graphs. Because of the particular choice of reference configuration, every plaquette has an odd number of links directed clockwise, and this is therefore a Kasteleyn graph \cite{Kasteleyn1961,Fisher1963}. In Figs.~\ref{FigRandomLoopsHeight}(b) and \ref{FigRandomLoopsHeightHc}(a), we show the loop mapping applied to typical configurations on the square and honeycomb lattices. Importantly, the choice of \(c_0\) also implies that each loop must span the vertical boundaries at least once. (The configurations shown in Figs.~\ref{FigRandomLoopsHeight} and \ref{FigRandomLoopsHeightHc} each consist of a single loop that spans the vertical boundaries multiple times and the horizontal boundaries once.)

The sum over dimer configurations \(c\) in \refeq{EqZah} is equivalent to a sum over loop configurations, which we also label \(c\). The dimer occupation number \(d_{\ell}\) is equal to the loop occupation number \(n_{\ell}\), except for those occupied in \(c_0\), for which \(d_{\ell} = 1 - n_{\ell}\). We can therefore use \refeq{EqScah2} to write the weighting in terms of a sum over link weights,
\beq[EqScahLoops]
S_c = S_{c_0} + 
\sum_{\ell} n_\ell (-1)^{\delta_{b_\ell,0}}
\zeta_\ell
\punc.
\eeq
where
\beq[EqDefinezeta]
\zeta_\ell = \ii a_\ell+ 2\vartheta_\ell^* \J_\ell + 2\vartheta_\ell \J_\ell^*
\punc,
\eeq
and similarly
\beq[EqScahLoops2]
\tv \cdot \Phiv_c = \tv \cdot \Phiv_{c_0} +
\sum_{\ell} n_\ell (-1)^{\delta_{b_\ell,0}}
\sum_\mu t_\mu\dsC^\mu_\ell
\punc,
\eeq
using \refeq{EqDefinePhi}.

The reference configuration \(c_0\) has \(d_{ib} = \delta_{b,0}\), and so its weighting is
\beq[EqScahRef]
S_{c_0} = \sum_{i} \left(\frac{1}{2}
\ii a_{i0}+ \ee^{+\ii \Qv\cdot \rv_i} J_{i} + \ee^{-\ii \Qv\cdot \rv_i}J_{i}^*
\right)\punc,
\eeq
using \refeq{EqScah} and \(\vartheta_{i0} = \ee^{-\ii \Qv\cdot \rv_i}\) from \refeq{EqVartheta}. From \refeq{EqDefinePhi2}, its flux is
\beq[EqPhivc0]
\big(\Phiv_{c_0}\big)_\mu = \frac{\Nuc}{L_\mu}\sum_{b} \left(\delta_{b,0} - \frac{1}{\goz}\right) \big(\uv_b\big)_\mu
\punc,
\eeq
where \(\Nuc = L_x L_y/\Auc\) is the number of unit cells. Because \(\sum_b \uv_b = \zerov\) and \(\uv_0 = -u\deltav_y\), we find \(\Phiv_{c_0} = -\Phi\sub{max}\deltav_y\), where
\beq[EqPhimax]
\Phi\sub{max} = \frac{L_x u}{\Auc} = \frac{L_x Q}{2\pi}
\punc,
\eeq
with \(Q\) defined as in \refsec{SecMagnetization}. The vertical flux \((\Phiv_{c_0})_y = -\Phi\sub{max}\) is therefore large and negative, scaling linearly with \(L_x\), and any other configuration has \((\Phiv_c)_y > -\Phi\sub{max}\).

\subsection{Grassmann integral for partition function}
\label{SecGrassmannIntegral}

We will now construct an exact expression for the partition function in terms of Grassmann integrals. First we define a pair of Grassmann variables \cite{Samuel1980}, \(\psi_i\) and \(\bar\psi_i\), on each site \(i\) of the lattice\footnote{An equivalent method, used in \refcite{Fendley2002} and \cite{Dijkgraaf2009}, defines a single Grassmann variable on each site and is closer to the original Pfaffian calculation; see \refapp{AppPfaffian}.} and let \(\Am\) be a weighted adjacency matrix for the (directed) Kasteleyn graph (i.e., \(A_{ij}\neq0\) if there an edge going from \(j\) to \(i\)), with weights to be specified below. We will relate the sum over loop configurations to the Grassmann integral
\beq
\int \prod_i \dd^2 \psi_i\, \ee^{-\bar\psiv(\idm - \Am)\psiv} = \det(\idm - \Am)
\punc,
\eeq
where \(\dd^2 \psi_i\) stands for \(\dd\bar\psi_i\dd\psi_i\) and \(\psiv\) and \(\bar\psiv\) are, respectively, column and row vectors with elements \(\psi_i\) and \(\bar\psi_i\).

To do so, we first show that expanding the exponential gives nonzero terms that correspond exactly to loop (and hence dimer) configurations, then show how each term can be given the correct sign, and finally specify the graph weights required to reproduce the configuration weightings in the partition function, \refeq{EqZah}.

\subsubsection{Loop configurations}
\label{SecLoopConfigurations}

Expanding \(\ee^{-\bar\psiv(\idm - \Am)\psiv}\) as a sum of monomials, the integral of a given term is nonzero only if, for every site \(i\), the Grassmann variables \(\psi_i\) and \(\bar\psi_i\) appear exactly once each. All such terms appear with a multiplicity that cancels the denominator in the Taylor expansion \cite{Ramirez2024}. If \(\psi_i\) and \(\bar\psi_i\) come from \(-\bar\psiv\idm\psiv = -\sum_i \bar\psi_i \psi_i\), then site \(i\) must appear nowhere else in this term; this corresponds to a site that is not connected to any loop. Otherwise, they come from \(\bar\psiv\Am\psiv\), which contains a term \(\bar\psi_i A_{ij} \psi_j\) if the Kasteleyn graph has an edge from \(j\) to \(i\). Each such site must therefore appear in two terms, once as the source and once as the target of the edge; a loop passes through this site in the direction specified by the Kasteleyn graph.

The sign of each term is determined by the convention
\beq[EqGrassmannIntegral]
\int \dd^2 \psi_i \, \psi_i \bar\psi_i = 1\punc.
\eeq
Sites \(i\) not connected to any loop contribute \(-\bar\psi_i\psi_i = \psi_i \bar\psi_i\), and therefore give \(+1\) on integration. A loop \(i_1 \rightarrow i_2 \rightarrow \dotsb \rightarrow i_n \rightarrow i_1\) gives a factor
\beq[EqGrassmannIntegralFactor]
\prod_{k=1}^{n}A_{i_{k+1},i_k}\int \prod_{k=1}^{n} \dd^2\psi_{i_k} \; \bar\psi_{i_1} \psi_{i_n} \dotsm \bar\psi_{i_3}\psi_{i_2} \bar\psi_{i_2} \psi_{i_1}
\punc.
\eeq
Moving \(\psi_{i_1}\) to the left of the integrand (anticommuting it with \(2n-1\) other Grassmann variables) and using \refeq{EqGrassmannIntegral} gives the integral as \(-1\), for a loop of any length. We therefore find
\beq[EqGrassmannIntegralLoops1]
\int \prod_i \dd^2 \psi_i\, \ee^{-\bar\psiv(\idm - \Am)\psiv} = \sum_c (-1)^{N_c}\prod_{\lambda \in c} \prod_{j\rightarrow i \in \lambda} A_{ij}
\punc,
\eeq
where \(N_c\) is the number of loops in configuration \(c\) and \(j\rightarrow i \in \lambda\) includes all directed edges from \(j\) to \(i\) in loop \(\lambda\).

\subsubsection{Loop sign correction}
\label{SecLoopSignCorrection}

For a dimer model with periodic boundary conditions, one can compensate for the minus sign associated with each loop by summing over fermion boundary conditions \cite{Kasteleyn1961,Lieb1967}. To do so, we set
\beq[EqWeightsA]
A_{ij} = (-1)^{\sum_\mu p_\mu \dsC^\mu_{j\rightarrow i}}A_{ij}'
\punc,
\eeq
where \(p_x\) and \(p_y\) are integers,\footnote{\(\pv = (\theta,\tau)\) in the notation of \refcite{Kenyon2006}, while \(p\) and \(\sigma\) of \refcite{Wilkins2021} are related to \(\pv\) by \(p = 1-p_x\) and \(\sigma = -(-1)^{p_y}\).} so that a link \(j \rightarrow i\) crossing the surface \(C^\mu\) (in either direction) has a sign \((-1)^{p_\mu}\). (The weights \(A_{ij}'\) will be specified in \refsec{SecLoopSourceWeighting}.) For each loop \(\lambda\), this gives
\beq
\prod_{j\rightarrow i \in \lambda} A_{ij} = 
(-1)^{w_x p_x+w_y p_y} \prod_{j\rightarrow i \in \lambda} A_{ij}'
\punc,
\eeq
where the winding \(w_\mu\) is the net number of times \(\lambda\) crosses \(C^\mu\) in the positive direction. [For example, the loop configuration shown in \reffig{FigRandomLoopsHeight}(b) has a single loop with \(\wv = (-1,2)\).] The directed nature of the loops implies that \(w_y \ge 1\). 

Each loop can be considered as a torus knot, a closed path on the surface of a torus, and a loop configuration \(c\) as a torus link. We use two properties of torus knots \cite{BurdeZieschang2003}: that the windings \(w_x\) and \(w_y\) are coprime\footnote{Note that \(0\) is coprime with \(\pm1\) but with no other integers, and so \(\wv = (0,1)\) is the only possibility with \(w_x = 0\).} and that, because any two loops do not intersect, their windings \(\wv\) and \(\wv'\) obey \(\epsilon_{\mu\nu}w_\mu w'_\nu = 0\). Together, these imply that all loops in a given configuration have equal windings, up to a sign; in fact, because \(w_y \ge 1\), the windings are equal. It follows that
\beq
\prod_{\lambda \in c} \prod_{j\rightarrow i \in \lambda} A_{ij} = 
(-1)^{N_c w_x p_x+N_c w_y p_y}\prod_{\lambda \in c} \prod_{j\rightarrow i \in \lambda} A_{ij}'
\eeq
for a configuration \(c\) containing \(N_c\) loops, each with winding \(\wv\).

We can now use the identity
\beq[EqpsSum]
\frac{1}{2}\sum_{p_x=0,1}\sum_{p_y=0,1} \varpi_{\pv} (-1)^{m_x p_x+m_y p_y} = -\varpi_{\mv}\punc,
\eeq
for any integers \(m_x\) and \(m_y\), where
\begin{align}
\label{Eqvarpi}
\varpi_{\pv} &= (-1)^{(1-p_x)(1-p_y)}\\
&= \begin{cases}
-1 & \text{if both \(p_x\) and \(p_y\) are even}\\
+1 & \text{otherwise,}
\end{cases}
\end{align}
to give
\beq
\frac{1}{2}\sum_{\pv}\varpi_{\pv} \prod_{\lambda \in c} \prod_{j\rightarrow i \in \lambda} A_{ij}
= - \varpi_{N_c \wv}
\prod_{\lambda \in c} \prod_{j\rightarrow i \in \lambda} A_{ij}'
\punc.
\eeq
The factor \(-\varpi_{N_c \wv}\) is \(+1\) if \(N_c w_x\) and \(N_c w_y\) are both even, and \(-1\) otherwise. Because \(w_x\) and \(w_y\), being coprime, cannot both be even, it is therefore equal to \((-1)^{N_c}\). This cancels the same factor in \refeq{EqGrassmannIntegralLoops1}, and so we have
\beq[EqGrassmannIntegralLoops2]
\frac{1}{2}\sum_{\pv}\varpi_{\pv}\int \prod_i \dd^2 \psi_i\, \ee^{-\bar\psiv(\idm - \Am)\psiv} = \sum_c \prod_{\lambda \in c} \prod_{j\rightarrow i \in \lambda} A_{ij}'
\punc.
\eeq

\subsubsection{Source weighting of loop configurations}
\label{SecLoopSourceWeighting}

Finally, to give each configuration \(c\) the appropriate weight in the partition function, we set \(A_{ij}'\) such that
\beq[EqSetAprime]
\prod_{\lambda \in c} \prod_{j\rightarrow i \in \lambda} A_{ij}' = \ee^{\ii \tv \cdot (\Phiv_c - \Phiv_{c_0})} \ee^{S_c-S_{c_0}}\punc.
\eeq
Together with \refeqand{EqScahLoops}{EqScahLoops2}, this implies
\beq
A_{ij}' = \exp \left[(-1)^{\delta_{b,0}}
\left(\zeta_{ib} + \ii \sum_\mu t_\mu \dsC_{ib}^\mu
\right)
\right]\punc,
\eeq
and hence, using \refeq{EqWeightsA},
\beq[EqAelement]
A_{ij} =  \exp \left[(-1)^{\delta_{b,0}}
\left(\zeta_{ib} - \ii \sum_\mu \phi_\mu \dsC_{ib}^\mu
\right)
\right]\punc,
\eeq
for sites \(i\) and \(j\) joined by a link \(j\rightarrow i\) along direction \(b\), where\footnote{The phases across the surfaces \(C^\mu\) depend only on this combination of \(\tv\) and \(\pv\) because each loop increases the flux \(\Phiv\) by \(\wv\), and so \(\Phiv_c = \Phiv_{c_0} + N_c \wv\).} \(\phiv = -\tv + \pi \pv\).

The nonzero elements in row \(i\) of the matrix \(\Am\), which take the values given in \refeq{EqAelement}, are determined by incoming edges of site \(i\). For \(i \in \mathrm{A}\), there is a single, vertical edge from \(\rv_j = \rv_i+\uv_0\), while a site \(i \in \mathrm{B}\) has incoming edges from \(\rv_j = \rv_i - \uv_b\) for \(b = 1, 2, \dotsc, \goz-1\), where vector arithmetic is modulo the periodic boundary conditions.

For explicit calculations, it is convenient to put the surfaces \(C^x\) and \(C^y\) at the (periodic) boundaries of the system. Specifically, we choose \(C^x\) just to the left of the line \(x = L_x\) and \(C^y\) just below \(y = L_y\) [as in \reffig{FigRandomLoopsHeight}(b) and \ref{FigRandomLoopsHeightHc}(a)], both oriented such that a link \(\ell\) crossing \(C^\mu\) in a positive Cartesian direction has \(\dsC^\mu_\ell = +1\). We can then write all elements of \(\Am\) as
\beq[EqAij]
A_{ij} = \begin{dcases}
\delta_{\rv_j,\rv_i+\uv_0}\super{\phiv} \ee^{-\zeta_{\ii 0}}
&\text{for \(i \in \mathrm{A}\)}\\
\sum_{b=1}^{\goz -1} \delta_{\rv_j+\uv_b,\rv_i}\super{\phiv} \ee^{\zeta_{\ii b}}
&\text{for \(i \in \mathrm{B}\),}
\end{dcases}
\eeq
where \(\delta\super{\phiv}\) is a Kronecker delta with boundary phases \(\phiv\),
\beq[EqDefinePeriodicKroneckerDelta]
\delta\super{\phiv}_{\rv,\rv'} = \begin{cases}
\ee^{- \ii \mv \cdot \phiv} & \text{if \(\rv = \rv' + \sum_\mu m_\mu \Lv_\mu\) for integer \(m_\mu\)}\\
0 & \text{otherwise.}
\end{cases}
\eeq

\subsubsection{Partition function}

We can now rewrite the partition function \(Z(\tv,a,\J)\) of \refeq{EqZah} explicitly in terms of Grassmann integrals. Using \refeqand{EqGrassmannIntegralLoops2}{EqSetAprime}, together with \(\Phiv_{c_0} = -\Phi\sub{max} \deltav_y\), we have
\beq[EqZahLoops]
Z(\tv,a,\J) =  \frac{1}{2}\sum_{\pv}\varpi_{\pv}
Z_{\phiv}(a,\J)
\punc,
\eeq
where \(\phiv = -\tv + \pi \pv\) and
\beq[EqZahphi]
Z_{\phiv}(a,\J) = \ee^{S_{c_0}(a,\J)}\ee^{ \ii \Phi\sub{max} (\phi_y - \pi p_y)}\int \prod_i \dd^2 \psi_i\, \ee^{-\bar\psiv(\idm - \Am)\psiv}\punc.
\eeq
The weighted adjacency matrix \(\Am\), defined in \refeq{EqAij}, implicitly depends on \(a\), \(\J\), and \(\phiv\) (and hence on \(\tv\) and \(\pv\)). The integral in \refeq{EqZahphi} is a Gaussian integral over Grassmann variables, and so can be evaluated as a determinant,
\beq[EqZahLoopsDet]
Z_{\phiv}(a,\J) = \ee^{S_{c_0}(a,\J)}\ee^{ \ii \Phi\sub{max} (\phi_y - \pi p_y)} \det (\idm - \Am)
\punc.
\eeq
These expressions are valid for any sources \(a\) and \(\J\), not necessarily slowly varying.

\subsection{Evaluation of partition function without sources}
\label{SecLatticeDeterminant}

As a first step towards an expression for the full partition function \(Z(\tv, a, \J)\), we evaluate it exactly with the sources \(a\) and \(\J\) both set to zero. This can be done by evaluating the determinant in \refeq{EqZahLoopsDet} and is equivalent to the standard calculation based on the Pfaffian \cite{Kasteleyn1961,Fisher1961}. We clarify the precise connection with the latter method in \refapp{AppPfaffian}.

Let \(\Am_0\) be the matrix defined in \refeq{EqAij} with \(\zeta_\ell = 0\). Up to the boundary phases \(\phiv\), it preserves translation symmetry with a unit cell containing one site of each sublattice. Its eigenvalues can therefore be labeled by wavevectors \(\kv \in \goB_{\phiv}\), the set of points in the corresponding Brillouin zone \(\goB\) such that
\beq[EqkSet]
k_\mu L_\mu \in 2\pi \dsZ + \phi_\mu
\punc.
\eeq
Using \refeq{EqDefinePeriodicKroneckerDelta}, these wavevectors obey
\beq[EqSumDeltaPhi]
\sum_{\rv} \delta\super{\phiv}_{\rv,\rv'} \ee^{\ii \kv \cdot \rv} = \ee^{\ii \kv \cdot \rv'}\punc,
\eeq
with any boundary phase canceled. Calculating the determinant of the \(2\times 2\) block for each \(\kv\), corresponding to the two sublattices, gives \(\det(\idm - \Am_0) = \prod_{\kv \in \goB_{\phiv}} P(\kv)\), where
\beq
P(\kv) = 
1 - \ee^{\ii \kv \cdot \uv_0}\sum_{b=1}^{\goz-1}\ee^{-\ii \kv \cdot \uv_b}\punc.
\eeq
When expressed in terms of the Bloch phases \(\ee^{\ii \kv \cdot \ev_1}\) and \(\ee^{\ii \kv \cdot \ev_2}\), where \(\ev_b = \uv_b - \uv_0\) is a lattice vector, this is exactly the characteristic polynomial for the dimer model \cite{Kenyon2006}, for a particular choice of Kasteleyn graph.\footnote{The same function is called \(L(p_x,p_y)\) in \refcite{Samuel1980}.}

The function \(P\) has zeroes at \(\kv_{\pm}\), where
\beq[EqDefinekpm]
\ee^{-\ii \kv_\pm \cdot \uv_b} = \begin{cases}
1 & \text{for \(b = 0\)}\\
-\omega^{\pm b} & \text{\phantom{for} \(b \ge 1\).}
\end{cases}
\eeq
For both square and honeycomb lattices, this is solved by \(\kv_{\pm} = \pm \frac{1}{2}\Qv\), with \(\Qv\) defined as in \refsec{SecMagnetization}. These are ``Dirac points'', where \(P(\kv_\pm + \qv) \sim \pm q_x + \ii q_y\) for small \(\lvert \qv\rvert\).

To evaluate the product over \(\kv\), we take the logarithm and then apply the Poisson summation formula. For wavevectors satisfying \refeq{EqkSet}, this can be expressed as\footnote{More explicitly,
\[
\sum_{k_\mu} f(k_\mu) = \int \dd k_\mu\, \frac{L_\mu}{2\pi} \Sha\left(\frac{k_\mu L_\mu - \phi_\mu}{2\pi}\right) f(k_\mu)
\punc,
\]
where \(\Sha(x) = \sum_{n=-\infty}^{n}\delta(x - n) = \sum_{m=-\infty}^\infty \ee^{2\pi \ii m x}\)
is a periodic delta function.}
\beq[EqPSFfk]
\sum_{k_\mu} f(k_\mu) = L_\mu \sum_{m_\mu = -\infty}^\infty \ee^{-\ii m_\mu \phi_\mu} \int \frac{\dd k_\mu}{2\pi} \ee^{\ii k_\mu m_\mu L_\mu} f(k_\mu)\punc,
\eeq
for any function \(f\), where the integral on the right-hand side is definite, over an interval containing all \(k_\mu\) values in the sum. This allows us to rewrite the determinant as
\beq[EqDetViaF]
\det(\idm - \Am_0) = \exp \left[ -\sum_{\mv \in \dsZ^2} \ee^{-\ii \mv \cdot \phiv} F(\mv)
\right]\punc,
\eeq
where
\beq[EqFIntegral]
F(\mv) = -L_x L_y \int_{\goB} \frac{\dd^2 \kv}{(2\pi)^2} \ee^{\ii (m_x k_x L_x + m_y k_y L_y)}\ln P(\kv)\punc.
\eeq

The term with \(\mv = \zerov\) is \(F(\zerov) = - L_x L_y \Sigma\), where
\beq
\Sigma = \int_{\goB} \frac{\dd^2 \kv}{(2\pi)^2}\ln P(\kv)
\eeq
is the entropy density (per unit area) in the thermodynamic limit, in agreement with standard results \cite{Samuel1980,Kenyon2006}. For the square lattice, the integral can be performed exactly to give \(\Sigma = G/\pi\) for unit lattice spacing, where \(G\) is Catalan's constant \cite{Kasteleyn1961,Fisher1961}.

For \(\mv \neq \zerov\) and large \(L_x\) and \(L_y\), the integral in \refeq{EqFIntegral} is dominated by the nonanalyticities of \(\ln P(\kv)\). These are at the Dirac points \(\kv_{\pm}\) (zeroes of \(P\)), as well as on the branch cut joining them, along which \(P\) is real and negative. For both lattices, we find the asymptotic large-size result
\beq[EqFresult]
F(\mv) \approx \delta_{m_x,0} \frac{\Phi\sub{max}}{m_y} + \frac{(-1)^{m_x\Phi\sub{max}}}{\pi (m_x^2/\rho + m_y^2\rho)}\punc,
\eeq
for \(\mv \neq \zerov\), where \(\rho = L_y/L_x\) and we have used \refeq{EqPhimax} for \(\Phi\sub{max}\). The first term comes from the branch cut, while the second is from using the linearized form of \(P(\kv_\pm + \qv)\) near the Dirac points. We can evaluate the contribution from the first term in \refeq{EqFresult} to the Fourier series in \refeq{EqDetViaF} using
\beq
\sum_{\substack{m_y = -\infty\\m_y \neq 0}}^{\infty}
\frac{\ee^{-\ii m_y \phi_y}}{m_y} = \ii (\phi_y-\pi)
\eeq
for \(0 < \phi_y < 2\pi\), giving a factor \(\ee^{-\ii \Phi\sub{max} (\phi_y-\pi)}\) in the determinant for all \(\phi_y\). It should noted that this \(\phiv\)-dependent contribution comes from all modes on the branch cut, including ones far from the Dirac points.

The result for the determinant is therefore
\begin{multline}
\label{EqDet1mAfinal}
\det(\idm - \Am_0) \approx \ee^{\Sigma L_x L_y} \ee^{-\ii \Phi\sub{max} \phi_y}\\
\times\exp \left( -\frac{1}{\pi}\sum_{\mv \neq \zerov} \frac{\ee^{-\ii \mv \cdot \phiv}(-1)^{m_x\Phi\sub{max}}}{ m_x^2/\rho + m_y^2\rho}
\right)\punc.
\end{multline}
Substituting this into \refeq{EqZahLoopsDet} and using \(S_{c_0}(0,0) = 0\) from \refeq{EqScahRef},
\begin{multline}
\label{EqZphi00}
Z_{\phiv}(0,0) \approx \ee^{\Sigma L_x L_y} (-1)^{\Phi\sub{max}(1-p_y)}\\
\times\exp \left( -\frac{1}{\pi}\sum_{\mv \neq \zerov} \frac{\ee^{-\ii \mv \cdot \phiv'}}{ m_x^2/\rho + m_y^2\rho}
\right)\punc,
\end{multline}
where \(\phiv' = \phiv + \deltav_x \pi \Phi\sub{max}\), and so the only remaining dependence on the reference configuration is through the parity of \(\Phi\sub{max}\).
Using an identity between Fourier series that we prove in \refapp{AppFourierSeriesIdentity}, this can be rewritten as
\begin{multline}
\label{EqZphi00b}
Z_{\phiv}(0,0) \approx \ee^{\Sigma L_x L_y} (-1)^{\Phi\sub{max}(1-p_y)}
\sqrt{\frac{\rho}{2}}[\eta(q)]^{-2} \\
\times\sum_{\Phiv\in\dsZ^2} (-1)^{\Phi_x\Phi\sub{max}} \omega_{\phiv}(\Phiv) \ee^{-\frac{\pi}{2}(\Phi_x^2/\rho + \Phi_y^2\rho)}
\punc,
\end{multline}
where \(\eta\) is the Dedekind eta function \cite[Sec.~23.15]{NIST:DLMF} and
\beq[EqDefineomegaphi]
\omega_{\phiv}(\Phiv) = -\ee^{-\ii \Phiv\cdot\phiv} \varpi_{\Phiv}
\punc,
\eeq
with \(\varpi\) defined in \refeq{Eqvarpi}.

The partition function \(Z(\tv,0,0)\) is given by \refeq{EqZahLoops}, with \(\phiv = -\tv + \pi \pv\). The sum over \(p_x\) and \(p_y\) can be evaluated (after a shift \(p_x \rightarrow 1-p_x\) for \(\Phi\sub{max}\) odd) using \refeq{EqpsSum}, which gives
\beq[EqpsSumomega]
\frac{1}{2}\sum_{\pv}\varpi_{\pv} \omega_{\phiv}(\Phiv) = \ee^{\ii \tv \cdot \Phiv}\punc,
\eeq
and hence the asymptotic large-size expression
\beq[EqZt00]
Z(\tv,0,0) \approx \ee^{\Sigma L_x L_y}
\sqrt{\frac{\rho}{2}}[\eta(q)]^{-2} \sum_{\Phiv\in\dsZ^2} \ee^{\ii \tv \cdot \Phiv} \ee^{-\frac{\pi}{2}(\Phi_x^2/\rho + \Phi_y^2\rho)}
\punc.
\eeq
Comparing with \refeq{EqZah}, the variate \(\Phiv\) can clearly be identified with the flux; equivalent results for its distribution have been derived on the square lattice using the transfer matrix \cite{Wilkins2021} and on the honeycomb lattice by the Pfaffian method \cite{Boutillier2009}.

\section{Dirac fermions}
\label{SecDiracFermions}

In this section, we start from the integral over lattice Grassmann variables \(\psi_j\) in \refeq{EqZahphi} and derive a corresponding integral over Grassmann-valued functions \(\psi(\rv)\) defined in continuous space. The resulting continuum theory should give the exact asymptotic behavior of any correlation function of the observables \(B_\ell\) and \(J_i\) at well-separated points, and hence must reproduce the partition function \(Z(\tv,a,\J)\). It is sufficient to restrict to slowly varying sources \(a\) and \(\J\), as discussed in \refsec{SecPartitionFunctionSources}, and also to omit ``contact terms'' such as \(a_\ell^2\), which cannot contribute to long-range correlations. We will furthermore argue that the continuum theory gives the correct asymptotic behavior not only for \(B_\ell\) and \(J_i\), but for all correlations of the dimer degrees of freedom at well-separated points.

The modes that dominate the long-distance behavior are those near the Dirac points \(\kv_\pm\) identified in \refsec{SecLatticeDeterminant}, where there are vanishing eigenvalues of \(\Am_0\), and so we construct the continuum theory by expanding around  \(\kv_\pm\). In doing so, we are in effect modifying modes with \(\kv\) far from the Dirac points, as well as adding an infinite set of additional modes beyond the lattice Brillouin zone. Neither of these can affect long-range correlation functions, because both theories are Gaussian and so different \(\kv\) modes are decoupled. They do, however, lead to a short-distance singularity in the fermion Green function, as expected for a continuum field theory. As discussed in \refsec{SecRegularizedContinuumDeterminant}, they also give a divergent contribution to the partition function, which requires regularization.

\subsection{Expansion around fermion zero modes}
\label{SecExpansion}

We first find an asymptotic expression for the action \(\bar\psiv (\idm - \Am)\psiv\) in \refeq{EqZahphi} for  \(\psiv\) ``close to'' a null vector of \(\idm - \Am_0\). More precisely, we allow \(\psiv\) to include only modes with \(\kv\) in a subset of those included in \refeq{EqkSet} that are near one of the Dirac points \(\kv_{\pm} = \pm\frac{1}{2}\Qv\). (Note that \(\kv_{\pm}\) themselves do not generically appear in this set.) Equivalently, in real space, we take the two null vectors, with elements \(\ee^{\ii \kv_{\pm} \cdot \rv_i}\) and equal amplitude on the two sublattices, and multiply by slowly varying functions of position.

Explicitly, we write \(\psiv \approx \psiv_+ + \psiv_-\) and \(\bar\psiv \approx \bar\psiv_+ + \bar\psiv_-\), in terms of vectors \(\psiv_{\pm}\) and \(\bar\psiv_{\pm}\) with components
\beq[Eqpsiexpansion]
\begin{aligned}
\psi_{j\pm} &= \sqrt{c}\,\ee^{\ii \kv_{\pm}\cdot \rv_j}\psi_\pm(\rv_j)\\
\bar\psi_{j\pm} &= \sqrt{c}\,\ee^{-\ii \kv_{\pm}\cdot \rv_j}\bar\psi_\pm(\rv_j)
\punc,
\end{aligned}
\eeq
respectively,\footnote{This can be compared to the corresponding operator expansion, Eq.~(36) of \refcite{Wilkins2023}.} where \(c\) is a positive constant to be chosen below and \(\psi_{\pm}(\rv)\) and \(\bar\psi_{\pm}(\rv)\) are slowly varying (Grassmann-valued) functions for \(0 \le r_\mu \le L_\mu\). These are constructed from the \(\kv\) modes specified above, for which \(\ee^{\ii \kv\cdot \Lv_\mu}= \ee^{\ii \phi_\mu}\), and hence obey twisted boundary conditions
\beq[EqpsipmPBCs]
\begin{aligned}
\psi_\pm(\rv + \Lv_\mu) &= \ee^{+\ii \phi_\mu} \psi_\pm(\rv)\\
\bar\psi_\pm(\rv + \Lv_\mu) &= \ee^{-\ii \phi_\mu} \bar\psi_\pm(\rv)\punc,
\end{aligned}
\eeq
as long as \(\Phi\sub{max}\) is even, which we assume in the following.\footnote{\label{FootnoteOddPhimax}Using \refeq{EqPhimax}, \(\ee^{\ii \kv_\pm \cdot \Lv_x} = \ee^{\ii \pi \Phi\sub{max}}\), and so if \(\Phi\sub{max}\) is odd, then \(\phiv\) in \refeq{EqpsipmPBCs} should be replaced by \(\phiv'\), defined after \refeq{EqZphi00}. This case can be handled by shifting \(p_x \rightarrow 1-p_x\), as in \refsec{SecLatticeDeterminant}.}

We will express \(\bar\psiv(\idm - \Am)\psiv\) in terms of the functions \(\psi_\pm\) and \(\bar\psi_\pm\), keeping terms up to first order in derivatives. As explained in \refsec{SecPartitionFunctionSources}, \(a_\ell\) and \(\J_i\) are treated as lattice discretizations of slowly varying continuum sources defined in \refeqand{EqDefineav}{EqDefinescJ}, and so we rewrite \refeq{EqDefinezeta} as
\beq[Eqzeta]
\zeta_{ib} = \ii \uv_b\cdot\av(\rv_i)+ \frac{1}{2}u\ee^{\ii \Qv \cdot \rv_i} \omega^{-b} \M(\rv_i) + \frac{1}{2}u\ee^{-\ii \Qv \cdot \rv_i} \omega^{b} \M^*(\rv_i)\punc,
\eeq
for \(i \in \mathrm{A}\), using \refeq{EqVartheta} for \(\vartheta_\ell\). As noted above, we will also drop contact terms involving \(\zeta_\ell^n\) for \(n \ge 2\).

We first use \refeq{EqAij} to calculate \((\idm - \Am)\psiv_{\pm}\). Due to their boundary conditions, \refeq{EqpsipmPBCs}, \(\psi_\pm\) and \(\bar\psi_\pm\) obey the same relation as \refeq{EqSumDeltaPhi}, giving
\begin{widetext}
\beq[EqAonpsi1]
[(\idm - \Am)\psiv_{\pm}]_i = 
\sqrt{c}\,\ee^{\ii \kv_{\pm}\cdot \rv_i}\times
\begin{dcases}
\psi_{\pm}(\rv_i)-\ee^{\ii \kv_{\pm}\cdot \uv_0}\ee^{- \zeta_{i0}}\psi_{\pm}(\rv_i+\uv_0)  
&\text{for \(i \in \mathrm{A}\)}\\
\psi_{\pm}(\rv_i)-\sum_{b=1}^{\goz -1}
\ee^{-\ii \kv_{\pm}\cdot \uv_b}\ee^{\zeta_{ib}}
\psi_{\pm}(\rv_i-\uv_b)
&\text{for \(i \in \mathrm{B}\).}
\end{dcases}
\eeq
We now expand \(\psi_\pm(\rv_i + \uv_0)\) and \(\psi_\pm(\rv_i - \uv_b)\) to linear order around \(\psi_\pm(\rv_i)\) and replace \(\ee^{\pm\zeta_{\ell}}\) by \(1 \pm \zeta_\ell\), to give
\beq[EqAonpsi2]
[(\idm - \Am)\psiv_{\pm}]_i \approx 
\sqrt{c}\,\ee^{\ii \kv_{\pm}\cdot \rv_i}\times
\begin{dcases}
\left(-\uv_0\cdot\del + \zeta_{i0}\right)\psi_{\pm}(\rv_i)
&\text{for \(i \in \mathrm{A}\)}\\
\sum_{b=1}^{\goz -1}\omega^{\pm b}
\left(-\uv_b\cdot\del + \zeta_{ib} \right)\psi_{\pm}(\rv_i)
&\text{for \(i \in \mathrm{B}\),}
\end{dcases}
\eeq
using \refeq{EqDefinekpm} for \(\ee^{-\ii \kv_{\pm}\cdot \uv_b}\).

Taking the inner product with \(\bar\psiv_{\pm}\), we find
\beq
\bar\psiv_{\pm}(\idm - \Am)\psiv_{\pm}
\approx c \sum_{i\in\mathrm{A}}
\bar\psi_\pm(\rv_i)
\left(-\uv_0\cdot\del + \zeta_{i0} \right)
\psi_{\pm}(\rv_i)
+
c \sum_{j\in\mathrm{B}}
\sum_{b=1}^{\goz -1}
\omega^{\pm b}
\bar\psi_\pm(\rv_j)
\left(-\uv_b\cdot\del + \zeta_{jb} \right)
\psi_{\pm}(\rv_j)\punc.
\eeq
\end{widetext}
The two terms can be combined by changing the site index in the second sum to \(i \in \mathrm{A}\) with \(\rv_i = \rv_j - \uv_b\), giving
\beq[EqpsiApsi1]
\bar\psiv_{\pm}(\idm - \Am)\psiv_{\pm}
\approx c \sum_{i\in\mathrm{A}}
\sum_{b=0}^{\goz -1}
\omega^{\pm b}
\bar\psi_\pm(\rv_i)
\left(-\uv_b\cdot\del + \zeta_{ib} \right)
\psi_{\pm}(\rv_i)
\punc,
\eeq
where we have used \(\psi_\pm(\rv_i + \uv_b) \approx \psi_\pm(\rv_i)\), and the same for \(\bar\psi_\pm\) and \(\del\psi_{\pm}\), and the fact that \(i,b\) and \(j,b\) refer to the same link.

If we now use the continuum expansion of \(\zeta_{ib}\), \refeq{Eqzeta}, then all functions of \(\rv_i\) appearing on the right-hand side are slowly varying on the lattice scale. The terms involving \(\M\), which contain rapidly oscillating factors \(\ee^{\pm \ii \Qv\cdot \rv_i}\), therefore sum to zero. The remaining terms can be calculated using
\beq[EqomegauSum]
\sum_{b=0}^{\goz -1}\omega^{\pm b}
\uv_b\cdot \vv
= -\frac{1}{2}u\goz v_{\mp}
\punc,
\eeq
where we define the complex scalar \(v_\pm = v_y \pm \ii v_x\) corresponding to any real vector \(\vv\).\footnote{To prove \refeq{EqomegauSum}, use the completeness relation, \refeq{EquuSum}, to write \(\frac{1}{2}\goz u^2 \vv = \sum_b (\uv_b \cdot \vv)\uv_b\) and then express the vectors on both sides as complex scalars using \(u_{b\pm}=-u\omega^{\mp b}\).} The result varies slowly in space, and so its sum can be approximated by an integral,
\beq[EqpsiApsiIntegral]
\bar\psiv_{\pm}(\idm - \Am)\psiv_{\pm} \approx
\frac{1}{2}c u \goz\int \frac{\dd^2 \rv}{\Auc}
\bar\psi_\pm(\rv)(\partial_{\mp} - \ii a_{\mp})\psi_{\pm}(\rv)
\punc,
\eeq
where \(\partial_\pm = \partial_y \pm \ii \partial_x\).

Similarly, taking the inner product of \refeq{EqAonpsi2} with \(\bar\psiv_{\mp}\) and using \(\ee^{\ii(\kv_+ - \kv_-)\cdot \rv_i} = \ee^{\ii \Qv \cdot \rv_i}\) gives
\begin{multline}
\bar\psiv_{\mp}(\idm - \Am)\psiv_{\pm}
\approx c\sum_{i\in\mathrm{A}} \ee^{\pm\ii \Qv \cdot \rv_i}
\sum_{b=0}^{\goz -1}
\omega^{\mp b}\\
\bar\psi_\mp(\rv_i)\left(-\uv_b\cdot\del + \zeta_{ib} \right)\psi_{\pm}(\rv_i)
\punc,
\end{multline}
where a factor of \(\ee^{\pm \ii \Qv \cdot \uv_b} = \omega^{\mp 2b}\) comes from changing the site index as in \refeq{EqpsiApsi1}. In this case, the oscillating factors mean that only the terms involving \(\M\) survive, and we find
\beq[EqpsiApsiIntegral2]
\bar\psiv_{\mp}(\idm - \Am)\psiv_{\pm}\approx \frac{1}{2}c u \goz\int \frac{\dd^2 \rv}{\Auc} \M_{\mp}(\rv)\bar\psi_\mp(\rv) \psi_\pm(\rv)
\punc,
\eeq
where \(\M_+(\rv) = \M(\rv)\) and \(\M_-(\rv) = \M^*(\rv)\).

Combining \refeqand{EqpsiApsiIntegral}{EqpsiApsiIntegral2} finally gives
\beq[EqFermionAction]
\bar\psiv(\idm - \Am)\psiv \approx
\int \dd^2\rv\, \scL[\psi,\bar\psi]\punc,
\eeq
with continuum action density
\begin{multline}
\label{EqFermionL1}
\scL[\psi,\bar\psi] = \bar\psi_+ (\partial_- - \ii a_-) \psi_+ + \bar\psi_- (\partial_+ - \ii a_+) \psi_-\\
+ \M_+ \bar\psi_+ \psi_- + \M_-\bar\psi_-\psi_+\punc,
\end{multline}
where we have chosen\footnote{The coefficient in \refeq{EqDefinescJ} was also chosen to simplify the coupling to \(\M_\pm\).} \(c = \frac{2\Auc}{u \goz}\) (equal to the distance \(\tau\) between plaquette centers). Apart from the inclusion of source terms, this action density is equivalent to the long-wavelength theory in terms of four Majorana fermion fields  found in \refcite{Fendley2002} and is the path-integral representation of the continuum Hamiltonian (with linearized dispersion) in \refcite{Wilkins2023}.

The action density can be rewritten in terms of the two-component column vector \(\psi = \begin{pmatrix}\psi_+&\psi_-\end{pmatrix}\tr\) and its Dirac conjugate \(\bar\psi = \begin{pmatrix}\bar\psi_+&\bar\psi_-\end{pmatrix}\gamma^y\), where we define the \(2\times 2\) Euclidean gamma matrices \(\gamma^\mu\) as equal to the Pauli matrices, \(\gamma^x = \sigma^x\) and \(\gamma^y = \sigma^y\). This gives \(\scL[\psi,\bar\psi] = \bar\psi (\slD + \rmM) \psi\), where
\beq
\rmM = \begin{pmatrix}
-\ii \M_-&0\\
0&\ii\M_+
\end{pmatrix}
\eeq
and \(\slD = \gamma^\mu (\partial_\mu - \ii a_\mu)\) is the Dirac operator. (Here and in the following, summation over repeated indices \(\mu\) and \(\nu\) is implied, unless stated otherwise.)

This is a theory of massless Dirac fermions in two ``space--time'' dimensions, in which the source term \(\av\) appears as a minimally coupled gauge field and \(\M\) is a mass that mixes the two fermion chiralities \(\psi_{\pm}\) \cite{Zinn-Justin2002}. In spite of the asymmetric reference configuration used to derive it, the continuum theory manifestly has rotational symmetry, which is in fact enlarged to continuous rotations. It also has gauge invariance, inherited from the microscopic theory, under \(a_\mu \rightarrow a_\mu + \partial_\mu \Chi(\rv)\) and \(\psi(\rv) \rightarrow \ee^{\ii \Chi(\rv)}\psi(\rv)\). Note that the square and honeycomb lattices have identical continuum theories, up to the multiplicative factors relating the microscopic and continuum sources.

The calculation leading to \refeq{EqFermionL1} justifies the claim, made in \refsec{SecPartitionFunctionSources}, that the sources \(\av\) and \(\M\) are the only ones required to describe the leading-order long-distance form of all dimer correlations. To see this, suppose terms with different dependence on \(\rv_i\) and \(b\) were added to \(\zeta_\ell\) in \refeq{Eqzeta}. Their contributions would sum to zero in both \refeqand{EqpsiApsiIntegral}{EqpsiApsiIntegral2}, because of the oscillating factors \(\omega^{\pm b}\) and \(\ee^{\pm \ii \Qv \cdot \rv_i}\), and they would therefore give no extra terms in the continuum action to this order. Note that the source terms that do survive are those that couple to bilinears of fermion modes at the Dirac points (i.e., with wavevector \(\zerov\) and \(\pm \Qv\)) and with the correct phase winding to match the chirality of these points.

\subsection{Regularized continuum determinant}
\label{SecRegularizedContinuumDeterminant}

The continuum Grassmann integral corresponding to the one in \refeq{EqZahphi} is therefore
\beq[EqIphiah0]
I_{\phiv}[\av,\M] = \int_{\phiv} \DD^2\psi \, \ee^{-\int \dd^2\rv \,\scL[\psi,\bar\psi]}
\punc,
\eeq
where the subscript \(\phiv\) indicates that the Grassmann fields satisfy the boundary conditions in \refeq{EqpsipmPBCs}. As argued above, this theory reproduces the correct dependence on \(\av\) and \(M\), because modes near the Dirac points are unmodified. On the other hand, modes away from these points also contribute to the partition function, and in particular to its dependence on \(\phiv\) (and hence \(\tv\)), and must be treated correctly to describe the flux distribution of the dimer model. In \refsec{SecLatticeDeterminant}, \(Z_{\phiv}(0,0)\) has been calculated exactly (for large system size) using the microscopic theory; here we will show that the continuum description can give a consistent result for \(I_{\phiv}[\zerov,0]\).

In fact, the continuum fermion integral in \refeq{EqIphiah0} is not well defined unless regularized. To do so, we write it as
\beq[EqZphiahDet]
I_{\phiv}[\av,\M] = \lim_{\Lambda\rightarrow \infty} \detLambda (\slD + \rmM)
\punc,
\eeq
with the regularized operator determinant \(\detLambda\) defined by
\beq[EqLnDetLambda]
\ln \detLambda (\slD + \rmM) = \Tr \left[\ee^{\slD^2/\Lambda^2} \ln (\slD + \rmM)\right]\punc,
\eeq
where \(\Tr\) denotes the trace in both matrix and operator senses, evaluated in the space of functions that obey \refeq{EqpsipmPBCs}. (The operator \(\slD^2\) has negative eigenvalues because \(\slD\) is antihermitian with these boundary conditions.) This regularization is gauge invariant, as required to preserve the gauge invariance of \(I_{\phiv}[\av,\M]\).

To calculate \(I_{\phiv}[\zerov,0]\), we require the eigenvalues of the matrix of operators \(\slashed\partial = \gamma^\mu\partial_\mu\). These are \(\pm \ii \lvert\kv \rvert\) with wavevectors \(\kv\) such that \(k_\mu L_\mu = 2\pi n_\mu + \phi_\mu\) (no sum over \(\mu\)), as in \refsec{SecLatticeDeterminant}, but in this case with \(n_\mu\) running over all integers. Using \(\ln (+\ii\lvert\kv\rvert) + \ln (-\ii\lvert\kv\rvert) = \ln (\lvert\kv\rvert^2)\) gives
\beq[EqLambdaRegularizedDet]
\ln \detLambda \slashed\partial
= \sum_{\kv} \ee^{-\lvert\kv\rvert^2/\Lambda^2} \ln (\lvert\kv\rvert^2)\punc,
\eeq
but it should be noted that this assumes a particular choice of branch cut for the logarithm. It therefore adds to \(\ln\detLambda \slashed\partial\) an imaginary number that is regularization dependent, and hence arbitrary, and may depend on the parameter \(\phiv\). This should be viewed as parallel to the appearance of a flux-dependent complex phase in the microscopic determinant, \refeq{EqDet1mAfinal}. The latter originates from modes along the branch cut which are far from the Dirac points, and so cannot be determined within the continuum model.

We can now use \refeq{EqPSFfk}, as in the lattice calculation but with the sum and integral extended to infinity, giving
\beq
\ln\detLambda \slashed\partial
= -\sum_{\mv \in \dsZ^2}\ee^{-\ii \mv \cdot \phiv} \tilde{F}_\Lambda(\mv)
\punc,
\eeq
where
\beq
\tilde{F}_\Lambda(\mv) = -L_x L_y
\int \frac{\dd^2 \kv}{(2\pi)^2}\ee^{\ii (m_x k_x L_x + m_y k_y L_y)} \ee^{-\lvert\kv\rvert^2/\Lambda^2} \ln (\lvert\kv\rvert^2)
\eeq
For nonzero \(\mv\), this integral is dominated by the logarithmic divergence at \(\kv = \zerov\), near which it agrees exactly with the corresponding integral for the microscopic determinant, \refeq{EqFIntegral}, near the Dirac points \(\kv_{\pm}\). The term with \(\mv = \zerov\) does not have a finite limit as \(\Lambda\rightarrow\infty\), and multiplies the determinant by a \(\phiv\)-independent constant. We therefore find
\beq[EqContinuumFreeFermionDeterminant]
I_{\phiv}[\zerov,0] = \detLambdainfty \slashed\partial \sim \exp \left( -\frac{1}{\pi}\sum_{\mv \neq \zerov} \frac{\ee^{-\ii \mv \cdot \phiv}}{ m_x^2/\rho + m_y^2\rho}
\right)\punc,
\eeq
where ``\(\sim\)'' means that this is an asymptotic expression for large system size, up to a factor that may depend on \(L_x\) and \(L_y\) but not \(\phiv\). Comparison with \refeq{EqZphi00} (see Footnote~\ref{FootnoteOddPhimax}) confirms that the continuum theory gives a result consistent with the microscopic model in the thermodynamic limit.

\subsection{Partition function in terms of Dirac fermions}

Finally, we write the full partition function \(Z(\tv,a,\J)\) in terms of the continuum fermionic model. To do so, we also need to find the weighting \(\ee^{S_{c_0}(\tv,a,\J)}\) of the reference configuration with \(a_\ell\) and \(\J_i\) treated as lattice discretizations of slowing varying continuum sources \(\av\) and \(\M\) according to \refeqand{EqDefineav}{EqDefinescJ}. Starting from \refeq{EqScahRef} and following the same logic as in \refsec{SecExpansion}, this gives
\beq
S_{c_0} \approx -\ii t_y \Phi\sub{max} + \frac{1}{2\Auc}\ii \uv_0 \cdot \int \dd^2 \rv\, 
\av(\rv)\punc.
\eeq
The integral vanishes using \refeq{EqaCondition2}, and so \(S_{c_0} \approx - \ii t_y \Phi\sub{max}\), to the same order as the approximations leading to \refeq{EqFermionAction}.

Combining these results with \refeqand{EqZahLoops}{EqZphi00}, we can therefore write the full partition function as
\beq[EqZahDirac]
Z(\tv,a,\J) \sim
\frac{1}{2}\sum_{\pv}\varpi_{\pv}
\int_{\phiv} \DD^2\psi \, \ee^{-\int \dd^2\rv\, \scL[\psi,\bar\psi]}
\punc,
\eeq
where \(\scL\) is given by \refeq{EqFermionL1} and the integral, over Grassmann-valued functions with the boundary conditions in \refeq{EqpsipmPBCs}, should be regularized as detailed above.

\section{Bosonic field theory}
\label{SecBosonization}

The fermionic theory in \refeq{EqZahDirac} is equivalent to one involving bosonic degrees of freedom, under a standard mapping referred to as bosonization \cite{vonDelft1998,Senechal2004}. In this section, we apply this mapping in the path-integral formulation \cite{Fujikawa2015} to find an expression for the partition function \(Z(\tv,a,\J)\) in terms of a Gaussian integral over a real field, which we interpret as the continuum height. All of the steps in this calculation can be found in the literature, but, for completeness, we give an outline here, using conventions appropriate for the dimer model, and details in the appendices. Throughout, we carefully account for the effects of finite system size and boundary conditions, which must be included correctly to describe the topological properties.

\subsection{Bosonization of continuum Grassmann integral}
\label{SecBosonization1}

Our goal is to find a bosonic representation of \refeq{EqZahDirac}, which, using \refeq{EqIphiah0}, can be written in terms of the regularized operator determinant in \refeq{EqZphiahDet}. We will do so by expressing this determinant in terms of expectation values of a real Gaussian field. Following \refcite{Fujikawa2015}, we first use a chiral gauge transformation to isolate the dependence on \(\av\), and then expand the remaining functional of \(\M\) in terms of \(2n\)-point correlations of fermionic bilinears.\footnote{Both steps effectively involving turning \(\av\) and \(\M\) on smoothly. This means that the ambiguity of the complex phase in the regularized determinant noted in \refsec{SecRegularizedContinuumDeterminant} is implicitly resolved by following the same branch of the logarithm.}

We will perform the calculation in the Lorenz gauge, \(\partial_\mu a_\mu = 0\), before restoring gauge invariance in the final result. With this choice, one can write \(a_\mu = \epsilon_{\mu\nu}\partial_\nu \beta\), where \(\beta(\rv)\) is smooth and periodic (see \refsec{SecPartitionFunctionSources}). Defining the chiral gamma matrix \(\gamma^z = -\ii \gamma^x \gamma^y = \sigma^z\), we have the operator identity
\beq[EqChiralGaugeIdentity]
\slD = \ee^{-\beta\gamma^z}\slashed{\partial}\ee^{-\beta\gamma^z}
\punc,
\eeq
using \(\gamma^\mu \gamma^z = -\gamma^z \gamma^\mu = -\ii \epsilon_{\mu\nu}\gamma^\nu\). The matrix--operator determinant in \refeq{EqZphiahDet} therefore obeys
\begin{align}
\frac{I_{\phiv}[\av,\M]}{I_{\phiv}[\av, 0]} &= \detLambda \left[
1_2 + \left(\ee^{-\beta \gamma^z}\slashed{\partial}\ee^{-\beta \gamma^z}\right)^{-1}
\rmM
\right]\\
&= \frac{I_{\phiv}[\zerov, \tilde\M]}{I_{\phiv}[\zerov, 0]}
\punc,
\label{EqIphiaM}
\end{align}
where \(\tilde\M_\pm = \M_\pm\ee^{\mp 2 \beta}\). To find \(I_{\phiv}[\av,\M]\), we require \(I_{\phiv}[\av,0]\) and \(I_{\phiv}[\zerov, \tilde\M]\), which we calculate in turn and then combine with \refeq{EqContinuumFreeFermionDeterminant} for \(I_{\phiv}[\zerov,0]\).

We start with \(I_{\phiv}[\av,0]\), expressed using \refeqand{EqZphiahDet}{EqLnDetLambda} as
\beq[EqIphiaStart]
\frac{I_{\phiv}[\av, 0]}{I_{\phiv}[\zerov,0]} = \lim_{\Lambda\rightarrow\infty}
\left[\Tr \ee^{\slD^2/\Lambda^2} \ln \slD - \Tr \ee^{\slashed{\partial}^2/\Lambda^2} \ln \slashed\partial\right]
\punc.
\eeq
In \refapp{AppDecouplingGaugeField}, we show that
\beq[EqIphiaResult]
\frac{I_{\phiv}[\av, 0]}{I_{\phiv}[\zerov,0]} = \exp \left[-\frac{1}{2\pi}\int \dd^2 \rv\, (\partial_\mu\beta) (\partial_\mu\beta)\right]
\punc,
\eeq
using \refeq{EqChiralGaugeIdentity}. This effectively amounts to applying a chiral gauge transformation \(\psi(\rv) \rightarrow \ee^{\beta\gamma^z} \psi(\rv)\) to the fermions, which gives a nonzero (``anomalous'') contribution due to the regulator \cite{Fujikawa1980,Bertlmann2000}.

Next, we calculate \(I_{\phiv}[\zerov, \M]\), by expanding in terms of correlations of fermionic bilinears, evaluating these exactly, and then resumming the series. The determinant is given by the Grassmann integral
\begin{multline}
\label{EqZphi0h}
I_{\phiv}[\zerov, \M] = \int_{\phiv} \DD^2\psi \, \ee^{-\int \dd^2 \rv \, (\bar\psi_+ \partial_- \psi_+ + \bar\psi_- \partial_+ \psi_-)}\\
\ee^{-\int \dd^2\rv \,
\M_+(\rv)\sigma_-(\rv)}
\ee^{-\int \dd^2\rv' \,
\M_-(\rv')\sigma_+(\rv')}\punc,
\end{multline}
where \(\sigma_\pm(\rv) = \bar{\psi}_\mp(\rv)\psi_\pm(\rv)\). Expanding the last two exponentials as power series gives
\begin{multline}
\label{EqZphi0h2}
\frac{I_{\phiv}[\zerov,\M]}{I_{\phiv}[\zerov, 0]} = \sum_{n=0}^{\infty}\sum_{n'=0}^\infty\frac{1}{n! n'!}\int \dd^2 \rv_1 \dotsm \dd^2 \rv_n \int \dd^2 \rv_1'\dotsm\dd^2\rv'_{n'} \\
\Delta_{\phiv}(\rv_1,\dotsc,\rv_n;\rv_1',\dotsc,\rv_n')\prod_{i=1}^n \M_+(\rv_i) \prod_{i'=1}^{n'} \M_-(\rv'_{i'})
\punc,
\end{multline}
where
\beq[EqDefineDelta]
\Delta_{\phiv}(\rv_1,\dotsc,\rv_n;\rv_1',\dotsc,\rv'_{n'}) = \left\langle
\prod_{i=1}^n \sigma_-(\rv_i)
\prod_{i'=1}^{n'}\sigma_+(\rv'_{i'})
\right\rangle_{\phiv}
\eeq
is an expectation value for free fermions with action density \(\scL = \bar\psi\slashed\partial\psi = \bar\psi_+ \partial_- \psi_+ + \bar\psi_- \partial_+ \psi_-\) and boundary phases \(\phiv\). Because the action is symmetric under independent phase rotations of the two chiralities, \(\psi_\pm \rightarrow \ee^{\ii c_\pm}\psi_\pm\), \(\bar\psi_\pm \rightarrow \ee^{-\ii c_\pm}\bar\psi_\pm\), this expectation value vanishes unless the number of \(\sigma_-\) and \(\sigma_+\) is equal, \(n = n'\).

In \refapp{AppBosonization} we show that \(\Delta_{\phiv}\) can be expressed exactly in terms of a real scalar field \(\h(\rv)\), as
\begin{multline}
\label{EqDeltaVertex}
\Delta_{\phiv}(\rv_1,\dotsc,\rv_n;\rv_1',\dotsc,\rv'_{n'}) \\=
\Lambda^{n+n'} 
\left\langle \prod_{i=1}^n \ee^{+2\pi \ii \h(\rv_i)} \prod_{i'=1}^{n'}\ee^{-2\pi \ii\h(\rv'_{i'})}\right\rangle_{\phiv}
\punc,
\end{multline}
where \(\Lambda^{-1}\) is a short-distance regularization parameter which can be chosen of order the lattice spacing. (We use the same symbol \(\Lambda\) as in the fermionic theory, but there is no need for them to be equal.) The expectation value is defined by
\beq[EqDefineVarphiEV]
\big\langle \mathcal{F}[\h] \big\rangle_{\phiv} = \frac{\sum_{\Phiv\in \dsZ^2} \int_{\Phiv}\DD\h(\rv)\, \mathcal{F}[\h]\, \ee^{-\scS\spr{b}_{\phiv}[\h]}}{\sum_{\Phiv\in \dsZ^2} \int_{\Phiv}\DD\h(\rv)\, \ee^{-\scS\spr{b}_{\phiv}[\h]}}
\punc,
\eeq
for any functional \(\mathcal{F}\), where
\beq[EqBosonAction]
\ee^{-\scS\spr{b}_{\phiv}[\h]} = \omega_{\phiv}(\Phiv)\exp \left[-\int \dd^2\rv\, \frac{\pi}{2}(\del\h)^2\right]
\punc,
\eeq
with \(\omega_{\phiv}\) defined as in \refeq{EqDefineomegaphi}. The integrals over \(\h\) in \refeq{EqDefineVarphiEV} can be defined by
\beq[EqhIntegralDecomposition]
\int_{\Phiv}\DD\h(\rv) \equiv \int_0^1 \dd h_0 \int \DD \xi(\rv)
\punc,
\eeq
with
\beq[EqhDecomposition]
h(\rv) = h_0 + \xi(\rv) -\sum_{\mu\nu}\frac{\epsilon_{\mu\nu} r_\mu\Phi_\nu}{L_\mu}
\punc,
\eeq
where \(\xi\) is a real field with periodic boundary conditions and zero spatial average.\footnote{In other words, the integral over \(\xi(\rv)\) in \refeq{EqhIntegralDecomposition} includes only configurations with \(\int \dd^2\rv \, \xi(\rv) = 0\). Because \(\ee^{-\scS\spr{b}_{\phiv}[\h]}\) is independent of the zero mode \(h_0\), isolating it in this way and restricting its range of integration is necessary to make integrals over \(h\) finite. (This is equivalent to treating \(h\) as compactified with radius \(1\).) As noted in \refapp{AppBosonization}, integrating over \(h_0\) gives \(\Delta_{\phiv}(\rv_1,\dotsc,\rv_n;\rv_1',\dotsc,\rv'_{n'}) = 0\) for \(n \neq n'\).} (The variate \(\Phiv\) will be identified with the flux below.)  With these definitions, the weighting \(\ee^{-\scS\spr{b}_{\phiv}[\h]}\) is independent of the zero mode \(h_0\) and factorizes into contributions from \(\xi\) and \(\Phiv\), which are therefore independently distributed. The variates \(\Phi_x\) and \(\Phi_y\) are integers with (complex) weight \(\omega_{\phiv}(\Phiv)\ee^{-\frac{\pi}{2}(\Phi_x^2/\rho + \Phi_y^2\rho)}\), while the field \(\xi(\rv)\) has action density \(\frac{\pi}{2}(\del\xi)^2\), up to regularization such that \(\Delta_{\phiv}\) is independent of \(\Lambda\) for separations much larger than \(\Lambda^{-1}\) (see \refsec{SecAppBosonicTheory}).

\begin{widetext}
Using \refeq{EqDeltaVertex}, the series in \refeq{EqZphi0h2} can be resummed to give
\beq
\frac{I_{\phiv}[\zerov, \M]}{I_{\phiv}[\zerov, 0]} =
\left\langle
\exp \int \dd^2\rv\left[
\Lambda\ee^{+2\pi \ii  \h(\rv)}\M_+(\rv)
+
\Lambda\ee^{-2\pi \ii  \h(\rv)} \M_-(\rv)
\right]
\right\rangle_{\phiv}
\punc.
\eeq
We now rewrite this expectation value using \refeq{EqDefineVarphiEV} and evaluate the denominator. The only \(\phiv\)-dependent contribution is from \(\Phiv\), which gives
\beq
\sum_{\Phiv\in\dsZ^2} \omega_{\phiv}(\Phiv)
\ee^{-\frac{\pi}{2}(\Phi_x^2/\rho + \Phi_y^2\rho)}
\punc.
\eeq
Using \refeqand{eq:defineQ}{eq:resultForQ1} in \refapp{AppFourierSeriesIdentity}, this is equal to \(I_{\phiv}[\zerov, 0]\) in \refeq{EqContinuumFreeFermionDeterminant} up to a constant factor, and so
\beq
I_{\phiv}[\zerov,\M]\propto 
\sum_{\Phiv\in \dsZ^2} \int_{\Phiv}\DD\h(\rv)\,
\ee^{-\scS\spr{b}_{\phiv}[\h]}
\exp \int \dd^2\rv\left[
\Lambda\ee^{+2\pi \ii  \h(\rv)}\M_+(\rv)
+
\Lambda\ee^{-2\pi \ii  \h(\rv)} \M_-(\rv)
\right]
\punc.
\eeq

We finally combine this with \refeqand{EqIphiaM}{EqIphiaResult} to write a bosonic representation of the full functional \(I_{\phiv}[\av,M]\). The result can be simplified with the transformation\footnote{The integration contour for \(\h(\rv)\) is shifted into the complex plane but can be deformed back to the real line without changing the value of the integral.} \(\h \rightarrow \h + \frac{1}{\ii \pi}\beta\) and the fact that
\beq
\ee^{-\scS\spr{b}_{\phiv}\left[\h + \frac{1}{\ii \pi}\beta\right]} = \ee^{-\scS\spr{b}_{\phiv}[\h]}\exp \int \dd^2\rv \left[ +\ii a_\mu \epsilon_{\mu\nu}\partial_\nu \h + \frac{1}{2\pi}(\partial_\mu\beta) (\partial_\mu\beta)\right]\punc,
\eeq
which follows from \refeq{EqBosonAction} and \(\delta_{\mu\rho} = -\epsilon_{\mu\nu}\epsilon_{\nu\rho}\). The result is
\beq[EqIphibosonic]
I_{\phiv}[\av,\M]\propto 
\sum_{\Phiv\in \dsZ^2} \int_{\Phiv}\DD\h(\rv)\,
\ee^{-\scS\spr{b}_{\phiv}[\h]}\\
\exp \int \dd^2\rv\left[
+\ii a_\mu \epsilon_{\mu\nu}\partial_\nu \h
+\Lambda\ee^{+2\pi \ii  \h(\rv)}\M_+(\rv)
+
\Lambda\ee^{-2\pi \ii  \h(\rv)} \M_-(\rv)
\right]
\punc.
\eeq
\end{widetext}
Though derived using the Lorenz gauge, this result is gauge invariant. To see this, consider a gauge transformation \(a_\mu \rightarrow a_\mu + \partial_\mu \Chi\), where \(\Chi\) has periodic boundary conditions; the term thus added to the exponential vanishes after integration by parts, because \refeq{EqhDecomposition} implies that \(\partial_\nu h\) is periodic.

The equivalence of the theories written in terms of fermions, \refeqand{EqFermionL1}{EqIphiah0}, and bosons, \refeq{EqIphibosonic}, can be expressed schematically as the bosonization identities \(\bar\psi\gamma^\mu\psi \equiv\epsilon_{\mu\nu}\partial_\nu \h\) and \(\psi_\mp(\rv)\bar\psi_\pm(\rv) \equiv \Lambda\ee^{\pm 2\pi \ii \h(\rv)}\) \cite{Fujikawa2015}. (The operator versions of these identities were used in \refcite{Wilkins2023}.)

\subsection{Continuum height theory}
\label{SecContinuumHeightTheory}

Finally, summing over \(p_x\) and \(p_y\) using \refeq{EqpsSumomega}, the full partition function, \refeq{EqZahDirac}, becomes
\beq[EqZahBosonic]
Z(\tv,a,\J) \sim
\sum_{\Phiv\in \dsZ^2} \ee^{\ii \tv \cdot \Phiv}
\int_{\Phiv}\DD\h(\rv)\,\ee^{-\scS\spr{b}[\h]} \ee^{S[\h,\av,\M]}
\punc,
\eeq
where the integral over \(h\) is defined by \refeq{EqhIntegralDecomposition}, which implies the boundary conditions
\beq[EqBosonicBCs]
\h(\rv + \Lv_\mu) - \h(\rv) = -\epsilon_{\mu\nu} \Phi_\nu
\punc.
\eeq
The action for the field \(h\) is
\begin{align}
\label{EqSbh1}
\scS\spr{b}[\h] &= \int \dd^2\rv\, \frac{\pi}{2}(\del\h)^2\\
&= \frac{\pi}{2}\left(\frac{1}{\rho}\Phi_x^2 + \rho\Phi_y^2\right) + \int \dd^2\rv\, \frac{\pi}{2}(\del\xi)^2
\punc,
\label{EqSbh2}
\end{align}
using \refeq{EqhDecomposition}. The coupling to the source terms is
\begin{multline}
\label{EqZahBosonicS}
S[\h,\av,\M] = 
\int \dd^2\rv\, \bigg[
\ii a_\mu \epsilon_{\mu\nu}\partial_\nu \h
+\Lambda\ee^{+2\pi \ii  \h(\rv)}\M(\rv)
\\
+\Lambda\ee^{-2\pi \ii  \h(\rv)} \M^*(\rv)\bigg]\punc,
\end{multline}
for microscopic sources \(a_\ell\) and \(\J_i\) are related to \(\av\) and \(\M\) by \refeqand{EqDefineav}{EqDefinescJ}. These expressions constitute our final result for the dimer partition function in terms of a real scalar field with zero mass and stiffness\footnote{We define the stiffness \(\kappa\) as the coefficient of \(\frac{1}{2}(\del h)^2\) in the action density. It is related to the coupling constant \(g\) for the Coulomb gas by \(\kappa = 2\pi g\) \cite{Nienhuis1987,Alet2006b}.} \(\kappa = \pi\), in exact agreement with the continuum height theory that is known to describe the dimer model \cite{Zeng1997,Fradkin2004,Alet2006b,Wilkins2023,Patil2014,Fradkin2013}.

Comparing \refeq{EqZahBosonicS} with \refeq{EqScahSmoothed} and using \(B_\mu(\rv) = \epsilon_{\mu\nu}\partial_\nu h(\rv)\), it is clear that \(h(\rv)\) is exactly the continuum version of the microscopic height \(h_p\), and that the coarse-grained magnetization, defined in \refeq{EqPsirv}, is given by \(\Psi(\rv)\approx \Lambda\ee^{-2\pi \ii h(\rv)}\). In introducing \(\h\) in \refeq{EqDeltaVertex} there is freedom to change the sign and shift \(h(\rv)\) by a constant (related to the spatial symmetries of the lattice model \cite{Alet2006b}), and we have chosen both to match the microscopic theory.

It is also apparent from its coupling to \(\tv\) in \refeq{EqZahBosonic} that \(\Phiv\) should be interpreted as the flux. The boundary condition on the continuum height, \refeq{EqBosonicBCs}, is identical to the microscopic boundary condition in \refeq{EqFluxTilt}.

\subsection{Dimers in terms of continuum fields}
\label{SecDimersViaContinuumFields}

As argued at the end of \refsec{SecExpansion}, the sources \(\av\) and \(M\) are the only ones required to describe long-distance correlations between dimers. It follows that these correlations are governed by the continuum fields \(\Bv(\rv)\) and \(\Psi(\rv)\), to which they are coupled in \refeq{EqZahBosonicS}. We can therefore express the dimer occupation \(d_\ell\) in terms of these fields such that it reproduces the leading asymptotic form of the correlations, by constructing the unique combination of \(\Bv\) and \(\Psi\) (and \(\Psi^*\)) consistent with the coarse-grained expressions in \refeqand{EqBvfromd}{EqPsifromd}. Using the completeness relation, \refeq{EquuSum}, and \(\Auc = \frac{1}{2}\tau u \goz\), this is given by
\begin{multline}
\label{EqDimersFromFields}
d_{ib} - \frac{1}{\goz} \approx \frac{\tau}{u} \uv_b \times \del h(\rv_i)
+\Lambda\tau
\ee^{-\ii \Qv\cdot \rv_i}
\omega^{+\eta_i b} \ee^{+2\pi \ii h(\rv_i)}
\\+\Lambda\tau
\ee^{+\ii \Qv\cdot \rv_i}
\omega^{-\eta_i b} \ee^{-2\pi \ii h(\rv_i)}
\punc.
\end{multline}

For the square lattice, in units where the nearest-neighbor distance is \(u = 1\), we have \(\tau = 1\), \(\omega = \ii\), and \(Q = \pi\). To express the dimer occupation number in a more conventional way, we refer the horizontal and vertical links with \(b = 1\) and \(2\) respectively to their A-sublattice site and those with \(b = 3\) and \(0\) respectively to their B-sublattice site. We can then rewrite \refeq{EqDimersFromFields} for this case as
\beq[EqDimersFromFieldsSq]
\begin{aligned}
d_{ix} - \frac{1}{4} &\approx (-1)^{x_i+y_i} \partial_y h(\rv_i) + \Lambda \left[ \ii (-1)^{x_i} \ee^{+2\pi \ii h(\rv_i)} + \text{c.c.}\right]\\
d_{iy} - \frac{1}{4} &\approx -(-1)^{x_i+y_i} \partial_x h(\rv_i) - \Lambda \left[ (-1)^{y_i} \ee^{+2\pi \ii h(\rv_i)}  + \text{c.c.} \right]
\punc,
\end{aligned}
\eeq
where \(d_{i\mu}\) is the occupation of the link between the sites at \(\rv_i\) and \(\rv_i + \deltav_\mu\). For the honeycomb lattice, where \(\omega = \ee^{2\pi \ii/3}\), we use units\footnote{The unit of length is the separation of two neighboring parallel dimers, as in \refcite{Fradkin2004}.} such that \(u = \frac{1}{\sqrt{3}}\), giving \(\tau = 1\) (see \reffig{FigVectors}) and \(Q = \frac{4\pi}{3}\). Referring every link to its A-sublattice site \(i\), the dimer occupation for the link between the sites at \(\rv_i\) and \(\rv_i + \uv_b\) (for \(b = 0, 1, 2\)) is
\beq[EqDimersFromFieldsHc]
d_{ib} - \frac{1}{3} \approx \hat{\nv}_b \cdot \del h(\rv_i) + \Lambda\left[ \ee^{-\frac{4\pi\ii}{3}x_i + \frac{2\pi\ii}{3}b}\ee^{+2\pi \ii h(\rv_i)} + \text{c.c.}\right]
\punc,
\eeq
where \(\hat{\nv}_b\) is a unit vector at an angle \(\frac{2\pi}{3}b\) counterclockwise from the \(+x\) direction. These expressions are equivalent to those stated previously on the square (e.g., \refcite{Fradkin2004,Papanikolaou2007,Wilkins2023}) and honeycomb (e.g., \refcite{Moessner2003,Fradkin2004,Patil2014}) lattices, up to scalings and uniform shifts of \(h(\rv)\) and different conventions for the short-distance regularization. (The continuum height field defined in \refcite{Wilkins2023} is \(\phi = -\frac{1}{4} - h\).)

These expressions, along with the decomposition in \refeq{EqhDecomposition}, allow one to find the asymptotic long-distance form of any correlation function in the original dimer model. For example, to find the two-point dimer--dimer correlation functions (with all sources set to zero), we require
\begin{align}
\left\langle \partial_\mu h(\rv) \, \partial_\nu h(\zerov) \right\rangle &= \left\langle \partial_\mu \xi(\rv) \, \partial_\nu \xi(\zerov) \right\rangle + \frac{\delta_{\mu\nu}}{L_\mu^2}\left\langle \lvert \Phiv \rvert^2 \right\rangle\\
\Lambda^2 \left\langle
\ee^{+2\pi \ii h(\rv)}\ee^{-2\pi \ii h(\zerov)}
\right\rangle &= W(\rv)\times \left\langle \ee^{-2\pi \ii\left( \frac{\Phi_y x}{L_x} - \frac{\Phi_x y}{L_y} \right)} \right \rangle
\label{EqhVertexCorrelator}
\end{align}
for \(\lvert\rv\rvert \gg \Lambda^{-1}\), where \(W(\rv) = \Lambda^2 \left\langle \ee^{+2\pi \ii [\xi(\rv) - \xi(\zerov)]}\right\rangle\). All expectation values are taken using the action \(\scS\spr{b}[\h]\) in \refeq{EqSbh2},\footnote{Note that \(\Omega_{\phiv}\) in \refeq{EqDefineOmega} is the expectation value of the same expression as on the right-hand side of \refeq{EqhVertexCorrelator}, but for a different distribution of \(\Phiv\).} which is translationally invariant. Any correlator that is not invariant under uniform shifts of \(h(\rv)\), such as \(\langle\ee^{+2\pi \ii h(\rv)}\ee^{+2\pi \ii h(\zerov)}\rangle\) or \(\langle\ee^{+2\pi \ii h(\rv)}\partial_\nu h(\zerov)\rangle\), vanishes due to the integral over \(h_0\) in \refeq{EqhIntegralDecomposition}.\footnote{This includes \(\left\langle \ee^{-2\pi \ii h(\rv)} \right\rangle = 0\), which is consistent with the corresponding statement \(\langle\ee^{-2\pi \ii h_i}\rangle = 0\) noted for the microscopic model in \refsec{SecMagnetization}.} The regularization parameter \(\Lambda\) is arbitrary and so must cancel in any correlation function.

Explicit expressions for \(W(\rv)\) and \(\langle \partial_\mu \xi(\rv) \, \partial_\nu \xi(\zerov) \rangle\), including finite-size effects, are given in \refeqand{EqDefineW}{EqdxiCorrelations} respectively. For \(\lvert\rv\rvert \ll L_x,L_y\), they simplify to
\begin{align}
\label{EqWasymptotic}
W(\rv) &\approx \frac{1}{(2\pi)^2\lvert \rv \rvert^2}\\
\langle \partial_\mu \xi(\rv) \partial_\nu \xi(\zerov) \rangle &\approx \frac{1}{2\pi^2}\frac{\lvert \rv \rvert^2 \delta_{\mu\nu} - 2r_\mu r_\nu}{\lvert \rv \rvert^4}\punc,
\label{EqdxiCorrnsAsymptotic}
\end{align}
and the contributions from \(\Phiv\) can be neglected. For separation \(\Rv = \rv_i - \rv_j\) much smaller than the system size, we therefore find
\begin{widetext}
\beq
\left\langle
\left(d_{i\mu} - \frac{1}{4}\right)
\left(d_{j\nu} - \frac{1}{4}\right)
\right\rangle
\approx \frac{1}{\pi^2 \lvert \Rv \rvert^4} \times
\begin{cases}
(-1)^{R_\mu} R_\mu^2 & \text{for \(\mu = \nu\) and \(R_{\perp\mu}\) even}\\
(-1)^{R_\mu} R_{\perp\mu}^2 & \text{for \(\mu = \nu\) and \(R_{\perp\mu}\) odd}\\
(-1)^{R_x + R_y} R_x R_y & \text{for \(\mu\neq\nu\)}
\end{cases}
\eeq
with \(R_{\perp\mu} = \epsilon_{\mu\rho}R_\rho\), on the square lattice using \refeq{EqDimersFromFieldsSq}, and
\beq
\left\langle
\left(d_{ib} - \frac{1}{4}\right)
\left(d_{jb'} - \frac{1}{4}\right)
\right\rangle
\approx \frac{(-1)^{R_x}}{2\pi^2 \lvert \Rv \rvert^4} 
\left[
\cos\left( \frac{4\pi}{3}R_x + \frac{2\pi}{3}(b-b')\right)
- \cos\left( 2\theta - \frac{2\pi}{3}(b+b')\right)
\right]
\eeq
\end{widetext}
with \(\theta = \tan^{-1}(R_y/R_x)\), on the honeycomb lattice using \refeq{EqDimersFromFieldsHc}. These agree with standard results for both lattices \cite{Fisher1963,Wilkins2023,Moessner2003}.

\section{Discussion and conclusions}

Starting from the classical dimer model on a bipartite lattice, we have derived a continuum field theory that reproduces the long-wavelength physics, including dimer correlations between well-separated links. The resulting theory, as well as the relationship with the microscopic dimer observables, agrees with the standard continuum height model that is known to provide the correct coarse-grained description \cite{Blote1982,Nienhuis1984,Zeng1997,Kenyon2001,Fradkin2004,Alet2006b,Fradkin2013,Patil2014}. We have included sources that couple to the flux density and magnetization (VBS order parameter), and shown explicitly that these are the only ones required to capture the continuum correlations. The continuum theory also describes the (discrete) flux distribution, which is related to the discontinuity of the height across the periodic boundaries.

The relationships given in \refsec{SecDimersViaContinuumFields} between the dimer occupations and the continuum height field allow arbitrary dimer--dimer correlations to be calculated between well-separated links, including the effects of boundary conditions. Similar relationships could be written in terms of the fermionic fields \(\psi(\rv)\) and \(\bar\psi(\rv)\), using the bosonization identities given at the end of \refsec{SecBosonization1}, but these must be averaged over boundary conditions (spin sectors) with appropriate (complex) weights, as in the partition function in \refeq{EqZahDirac}.

The continuum height theory is certainly not new, but in previous works it is typically justified as a heuristic coarse-graining of the microscopic height, with structure dictated by general symmetry and topology considerations and with coefficients fixed by matching correlations to the exact solution. The continuum height was rigorously shown to be distributed as a massless free field for closed boundary conditions in \refcite{Kenyon2001}. Continuum fermionic theories related to the one derived in \refsec{SecDiracFermions} were presented in \refcite{Fendley2002} and \cite{Giuliani2015} in terms of Majorana modes (see \refsec{SecExpansionMajorana}) and \refcite{Dijkgraaf2009}, where it is shown that the Kasteleyn graph provides a discretization of the Dirac operator. By including source terms, the approach used here leads not only to the correlations of the height, as in \refcite{Kenyon2001} and \cite{Giuliani2015}, but also to the leading-order expressions for the dimer occupation in terms of the height.

In \refcite{Wilkins2023}, the same continuum theory was derived starting from the transfer matrix written in terms of fermion operators and using the operator form of the bosonization identities. The equivalence of the approaches can be understood by interpreting the directed loops defined in \refsec{SecGrassmannIntegrals} as the world-lines of fermionic particles in imaginary time. More precisely, the operator trace can be rewritten in terms of Grassmann integrals by applying the usual path-integral mapping via fermion coherent states. A na\"\i ve application of this approach gives Grassmann variables on the links of lattice, but one can move these to the sites by introducing additional Grassmann integrals, as shown for the honeycomb lattice in \refcite{Smerald2018}.

The approach used here, based on the real-space picture, can be considered more natural than the transfer-matrix method and straightforwardly allows the square and honeycomb lattices to be treated on an equal footing. The stiffness \(\kappa=\pi\) in the effective action, \refeq{EqSbh1}, is the same for both, and would in fact be the same for any isotropic model described by free Dirac fermions in the continuum. Anisotropic weights in the original dimer model give anisotropic stiffness \cite{Wilkins2023}, equivalent to rescaling the spatial coordinates.

Because the full spatial structure is maintained until the continuum limit is taken, this approach allows more general modifications such as applying arbitrary link potentials.\footnote{In \refsec{SecExpansion} the replacement \(\ee^{\zeta_\ell} \approx 1 + \zeta_\ell\) is used to simplify the link weighting, because higher terms do not contribute to correlations between links. This cannot be done for general potentials, meaning that there is not the same direct correspondence between microscopic potentials and terms in the continuum action.} Provided that the Dirac points in the spectrum remain, the resulting continuum theory would still involve free Dirac fermions, and hence a free bosonic field with \(\kappa = \pi\); a modification that opens a gap would instead result in a trivial long-distance theory. (This can be the case even without potentials, as with the square--octagon lattice \cite{Salinas1974}.) Interactions between dimers, which additionally give cosine terms that drive ordering transitions \cite{Alet2006b}, can in some cases can be handled perturbatively \cite{Falco2013,Wilkins2023,Giuliani2015,Giuliani2017}. It may also be possible to extend the method to cases without translation symmetry, such as quasicrystals \cite{Flicker2020,Singh2024} and models with spatial disorder \cite{Caracciolo2021}.

Defects in the close-packing constraint, referred to as monomers, are known to correspond to topological defects in the height-field theory \cite{Nienhuis1987,Fendley2002,Allegra2015,Wilkins2023}. The partition function with inserted monomers can be expressed in terms of correlations of a vertex operator dual to \(e^{\pm 2\pi \ii h}\), but we leave a similar constructive derivation of this relationship to future work.

\acknowledgments

I would like to thank Neil Wilkins for collaborations on related work. I am grateful to Paul Saffin for helpful discussions.

\section*{Appendices}

\appendix

\section{Connection to solution via Pfaffian}
\label{AppPfaffian}

In \refsec{SecGrassmannIntegral}, the partition function of the dimer model is expressed in terms of an integral over a pair of Grassmann variables \(\psi_i\) and \(\bar\psi_i\) on each site \(i\), which evaluates to a determinant. Here we briefly review an alternative construction based on a single Grassmann variable \(\psi_i\) and a corresponding Pfaffian. The two methods are equivalent and differ only in the details, but the Pfaffian method is more convenient for closed boundary conditions.

\subsection{Grassmann integral}

The dimer partition function can be written as the Pfaffian of the Kasteleyn matrix \(\Km\) \cite{Kasteleyn1961,Fisher1963}, which is equal to an integral over a single Grassmann variable \(\psi_i\) on each site \cite{Fendley2002,Dijkgraaf2009}
\beq[EqPfaffianZ1]
\Pf \Km = \int \DD\psiv \, \ee^{-\frac{1}{2}\psiv\tr\Km \psiv}
\punc.
\eeq
The matrix \(\Km\) is the antisymmetrized directed adjacency matrix for the Kasteleyn graph, with weights specified below.

With closed boundary conditions (a planar graph), any choice of Kasteleyn graph, defined such that every plaquette has an odd number of links directed clockwise \cite{Kasteleyn1961,Fisher1963}, gives the correct relative sign for each configuration. To allow for periodic boundary conditions, one can fix the order of the integrals (and hence the sign of the Pfaffian) by defining
\beq
\DD \psiv = \prod_{j\rightarrow i \in c_0} \dd\psi_j \dd \psi_i
\eeq
for a reference configuration \(c_0\). The notation \(j \rightarrow i \in c\) means that the link between sites \(i\) and \(j\) is occupied in configuration \(c\) and that the corresponding edge in the Kasteleyn graph is directed from \(j\) to \(i\). Expanding the exponential in \refeq{EqPfaffianZ1} then gives
\begin{align}
\Pf \Km &= \sum_c \int \prod_{j\rightarrow i \in c_0} \dd\psi_j \dd \psi_i \prod_{j\rightarrow i \in c} \psi_{i}\psi_{j}K_{ij}\\
&= \sum_c \prod_{j\rightarrow i\in c}K_{ij} \int \prod_{j\rightarrow i \in c_0 \setminus c}  \dd\psi_j \dd \psi_i \prod_{j \rightarrow i \in c \setminus c_0} \psi_i \psi_j
\punc,
\end{align}
where \(c_1\setminus c_2\) is the set of links occupied in configuration \(c_1\) but not in \(c_2\).

For the reference configuration and associated Kasteleyn graph specified in \refsec{SecDirectedLoops}, the Grassmann integrals can be interpreted in terms of the same directed loop configurations as in \refsec{SecLoopConfigurations}. In this case, a loop \(i_1 \rightarrow i_2 \rightarrow \dotsb \rightarrow i_n \rightarrow i_1\) with \(i_1 \rightarrow i_2 \in c_0\) gives a factor
\begin{multline}
\prod_{k=1}^{n}K_{i_{k+1},i_k}\int \dd\psi_{i_1}\dd\psi_{i_2}\dotsm\dd\psi_{i_{n-1}}\dd\psi_{i_n} \\\psi_{i_3}\psi_{i_2}\psi_{i_5}\psi_{i_4}\dotsm \psi_{i_{n-1}}\psi_{i_{n-2}}\psi_{i_1}\psi_{i_n}
\punc,
\end{multline}
analogous to \refeq{EqGrassmannIntegralFactor}. Rearranging the Grassmann variables (and using the fact that \(n\) is even for any bipartite lattice), gives the integral as \(-1\), and so we find
\beq
\Pf \Km = \sum_c (-1)^{N_c}\prod_{j\rightarrow i \in c} K_{ij}
\punc,
\eeq
analogous to \refeq{EqGrassmannIntegralLoops1}.

To give appropriate weights to each configuration (including the boundary phases as in \refsec{SecLoopSignCorrection}), the elements of \(\Km\) should be chosen as
\beq
K_{ij} = -K_{ji} = \sum_{b} (-1)^{\delta_{b,0}}\delta_{\rv_i+\uv_b,\rv_j}\super{\phiv} \ee^{\zeta_{\ii b}}
\eeq
for \(i \in \mathrm{A}\), which implies that \(K_{ij} = A_{ij}\) for \(j\rightarrow i \not\in c_0\) and \(K_{ij} = A_{ij}^{-1}\) for \(j \rightarrow i \in c_0\). With this choice, we find
\beq
\Pf \Km = \prod_{j\rightarrow i \in c_0}A_{ij}^{-1} \times \sum_c (-1)^{N_c} \prod_{j\rightarrow i \in c\triangle c_0} A_{ij}\punc,
\eeq
where \(c \triangle c' = (c \setminus c')\cup(c' \setminus c)\) is the symmetric difference. Using \refeqand{EqGrassmannIntegralLoops1}{EqAelement}, then gives
\beq
\Pf \Km = \ee^{S_{c_0}-\ii \phiv\cdot \Phi_{c_0}}\det (\idm - \Am)\punc,
\eeq
relating the Pfaffian to the determinant constructed in the main text.

\subsection{Expansion around Dirac points}
\label{SecExpansionMajorana}

A continuum theory can be derived starting from \refeq{EqPfaffianZ1} in the same way as in \refsec{SecExpansion}. Here, we do so omitting source terms and neglecting the boundary conditions for simplicity.

For \(\zeta_\ell = 0\), the matrix \(\Km\) has null vectors at the same wavevectors \(\kv_{\pm}\) as \(\idm - \Am\). Using the bipartite nature of the lattice, we can expand in terms of two continuum fields for each sublattice, \(\psi_{\mathrm{A}\pm}\) and \(\psi_{\mathrm{B}\pm}\),
\beq
\psi_i \approx \sqrt{c}\left[
\ee^{\ii \kv_+ \cdot \rv_i} \psi_{s+}(\rv_i)+\ee^{\ii \kv_- \cdot \rv_i} \psi_{s-}(\rv_i)
\right]
\punc,
\eeq
for \(i \in s = \mathrm{A}\) or \(\mathrm{B}\). Following the same steps as leading to \refeq{EqFermionL1}, one finds
\beq
\frac{1}{2}\psiv\tr\Km\psiv \approx \int \dd^2 \rv\, \big(\psi_{\mathrm{B}-}\partial_- \psi_{\mathrm{A}+} + \psi_{\mathrm{B}+}\partial_+ \psi_{\mathrm{A}-}\big)
\punc,
\eeq
with the same choice of the constant \(c\). This is equivalent to the theory derived in \refsec{SecExpansion}, with the identification \(\psi_{\mathrm{A}\pm} = \psi_{\pm}\) and \(\psi_{\mathrm{B}\pm} = \bar\psi_{\mp}\), which amounts to a particle--hole transformation on one sublattice.

One can instead take even and odd linear combinations of the fields on the two sublattices, giving a theory with four Majorana modes \cite{Fendley2002}. On the square lattice, these modes correspond to points with \(k_y = 0\) and \(\pi\) in the Brillouin zone for the lattice unit cell containing one site.

\section{Fourier series exponential identity}
\label{AppFourierSeriesIdentity}

\newcommand{\Qt}{\tilde{Q}}

We define
\beq[eq:defineQ]
Q(\phiv) = \exp\left[-\sum_{m_x = -\infty}^\infty\sum_{m_y = -\infty}^\infty \ee^{-\ii \mv \cdot \phiv} \Qt(m_x,m_y)\right]
\punc,
\eeq
where \(\Qt(0,0) = 0\) and
\beq
\Qt(m_x,m_y) = \frac{1}{\pi}\frac{1}{m_x^2/\rho + m_y^2\rho}
\eeq
otherwise. We will show that
\begin{align}
Q(\phiv) &= q^{2\left(\frac{\phi_y}{2\pi}\right)^2}
\left\lvert\frac{\theta_1\!\left(\frac{1}{2}\varphi, q\right)}{\eta(q)}\right\rvert^2
\label{eq:resultForQ2}\\
&= -\sqrt{\frac{\rho}{2}}[\eta(q)]^{-2} \sum_{\mv} \ee^{-\ii \mv \cdot \phiv} \varpi_{\mv} \ee^{-\frac{\pi}{2}(m_x^2/\rho + m_y^2\rho)}
\label{eq:resultForQ1}
\punc,
\end{align}
where \(\varphi = \phi_x + \ii \phi_y/\rho\), \(q = \ee^{-\pi/\rho}\), \(\theta_1\) is an elliptic theta function \cite[Ch.~20]{NIST:DLMF}, \(\eta\) is the Dedekind eta function \cite[Sec.~23.15]{NIST:DLMF}, and \(\varpi\) is defined in \refeq{Eqvarpi}.

These identities are used in the main text to rewrite \refeq{EqZphi00}, which arises from a sum over fermion modes near the Dirac points, as \refeq{EqZphi00b}, interpreted as a sum over flux sectors, and also to relate the fermionic partition function in \refeq{EqContinuumFreeFermionDeterminant} to the bosonic partition function in the denominator of \refeq{EqDefineVarphiEV}. They are also used to simplify the sum over flux sectors in \refsec{SecAppBosonicTheory} and to calculate the bosonic Green function in \refsec{AppBosonGreenFunction}, based on the fact that \refeq{eq:defineQ} implies
\beq[EqQpde]
\pi\left(\frac{1}{\rho}\frac{\partial^2}{\partial \phi_x^2} + \rho\frac{\partial^2}{\partial \phi_y^2}\right) \ln Q(\phiv) = \Sha\!\left(\frac{\phi_x}{2\pi}\right)\Sha\!\left(\frac{\phi_y}{2\pi}\right) - 1
\eeq
where \(\Sha(x) = \sum_{n=-\infty}^{n}\delta(x - n) = \sum_{m=-\infty}^\infty \ee^{-2\pi \ii m x}\)
is a periodic delta function
(and the constant is due to the omitted \(\mv = \zerov\) mode).

We start by performing the sum over \(m_y\). For \(m_x = 0\), this gives
\begin{align}
\sum_{\substack{m_y = -\infty\\m_y \neq 0}}^{\infty} \Qt(0,m_y) \ee^{-\ii m_y \phi_y} &= \frac{1}{\rho\pi}\sum_{\substack{m_y = -\infty\\m_y \neq 0}}^\infty \frac{1}{m_y^2}\ee^{-\ii m_y \phi_y}\\
&= \frac{1}{\rho}\left[\frac{(\phi_y - \pi)^2}{2\pi} - \frac{\pi}{6}\right]
\label{eq:sumovermy1}
\punc,
\end{align}
if \(0 \le \phi_y \le 2\pi\) (which we assume in the following, before restoring periodicity in the final result). For \(m_x \neq 0\), we use the Poisson summation formula,
\beq[eq:PSF]
\sum_{m = -\infty}^{\infty} f(m) = \sum_{n=-\infty}^{\infty}\int_{-\infty}^\infty \dd m \, \ee^{2\pi \ii m n} f(m)\punc,
\eeq
to give
\begin{widetext}
\begin{align}
\sum_{m_y=-\infty}^{\infty}\Qt(m_x,m_y)\ee^{-\ii (m_x \phi_x + m_y \phi_y)} &= \frac{\ee^{-\ii m_x \phi_x}}{\pi}\sum_{n=-\infty}^{\infty} \int_{-\infty}^{\infty}\dd m_y \frac{\ee^{-\ii m_y \left(\phi_y - 2\pi n\right)}}{m_x^2/\rho + m_y^2\rho}\\
&= \frac{\ee^{-\ii m_x \phi_x}}{\lvert m_x \rvert}\sum_{n=-\infty}^{\infty}\ee^{-\left\lvert \phi_y - 2\pi n\right\rvert \lvert m_x \rvert / \rho}
\punc,
\end{align}
where we have again assumed \(0 \le \phi_y \le 2\pi\).

Using the Taylor expansion of the logarithm,
\beq
\sum_{m = 1}^\infty \frac{w^m}{m} = -\ln(1-w)
\eeq
for \(\lvert w \rvert < 1\), the sum over nonzero \(m_x\) is
\beq
\sum_{\substack{m_x = -\infty\\m_x \neq 0}}^{\infty}\sum_{m_y=-\infty}^{\infty} \Qt(m_x,m_y)\ee^{-\ii (m_x \phi_x + m_y \phi_y)} = -\sum_{n=-\infty}^\infty \left[\ln\left(1 - \ee^{-\ii \phi_x}\ee^{-\lvert\phi_y - 2\pi n\rvert / \rho}\right) + \text{c.c.}\right]
\punc,
\eeq
Substituting this and \refeq{eq:sumovermy1} into \refeq{eq:defineQ} gives
\beq
Q(\phiv) = q^{2\left(\frac{\phi_y}{2\pi}\right)^2} q^{1/3} \ee^{\phi_y/\rho} \left\lvert\prod_{n=-\infty}^{\infty}
\left(1 - \ee^{-\ii \phi_x}\ee^{-\lvert\phi_y - 2\pi n\rvert / \rho}\right)\right\rvert^2
\punc,
\eeq
where \(q = \ee^{-\pi/\rho}\). Using the infinite-product definitions of the elliptic theta function \(\theta_1\) \cite[Eq.~(20.5.1)]{NIST:DLMF} and the Dedekind eta function \(\eta\) \cite[Eq.~(23.17.8)]{NIST:DLMF}, this can be written as \refeq{eq:resultForQ2}, which is periodic (due to the quasiperiodicity of \(\theta_1\)), and so applies for all \(\phi_y\).

Writing \(\theta_1\) and its conjugate as Fourier series \cite[Eq.~(20.2.1)]{NIST:DLMF} gives
\beq
Q(\phiv) = [\eta(q)]^{-2} q^{2\left(\frac{\phi_y}{2\pi}\right)^2} \sum_{m = -\infty}^{\infty}\sum_{n = -\infty}^{\infty} (-1)^{m+n} q^{(m+\frac{1}{2})^2+(n+\frac{1}{2})^2}\ee^{-(m+n+1)\phi_y/\rho}\ee^{-\ii \phi_x (m-n)}
\punc.
\eeq
Replacing the sum over \(m\) by one over \(m_x = m - n\) and applying \refeq{eq:PSF} to the sum over \(n\) gives
\begin{equation}
Q(\phiv) = [\eta(q)]^{-2} q^{2\left(\frac{\phi_y}{2\pi}\right)^2} \sum_{m_x = -\infty}^{\infty} \sum_{m_y = -\infty}^{\infty} \ee^{-\ii m_x \phi_x} (-1)^{m_x} q^{m_x^2 + m_x + \frac{1}{2}} q^{(m_x+1)\frac{\phi_y}{\pi}}
\int_{-\infty}^{+\infty}\dd n\,\ee^{2\pi \ii m_y n} q^{2n^2} q^{2n\left(m_x+1+\frac{\phi_y}{\pi}\right)}
\punc,
\end{equation}
which evaluates to \refeq{eq:resultForQ1}.
\end{widetext}

\section{Gauge field contribution to continuum determinant}
\label{AppDecouplingGaugeField}

In this appendix, we show that the Grassmann integral \(I_{\phiv}[\av, 0]\), given by \refeq{EqIphiaStart}, evaluates to \refeq{EqIphiaResult} in the Lorenz gauge, \(a_\mu(\rv) = \epsilon_{\mu\nu}\partial_\nu \beta(\rv)\). This calculation is equivalent to finding the Jacobian for a chiral gauge transformation of the Grassmann integral \cite{Roskies1981,Fujikawa2015,Bertlmann2000}, but here we apply it directly to the regularized determinant.

We start by defining \(\slD_q = \gamma^\mu ( \partial_\mu - \ii q a_\mu)\), the Dirac operator with charge \(q\), and writing
\beq[EqIntegralOverq]
\frac{I_{\phiv}[\av, 0]}{I_{\phiv}[\zerov,0]} = \lim_{\Lambda\rightarrow\infty} \int_0^1 \dd q\, \frac{\dd}{\dd q} \Tr \ee^{\slD_q^2/\Lambda^2} \ln \slD_q
\punc.
\eeq
Because \(\slD_q^\dagger = -\slD_q\) and \(\gamma^z \slD_q = -\slD_q \gamma^z\), the eigenvalues of \(\slD_q\) are imaginary and come in complex conjugate pairs. We therefore pair up these eigenvalues, as in \refeq{EqLambdaRegularizedDet}, giving
\begin{align}
\Tr \ee^{\slD_q^2/\Lambda^2} \ln \slD_q &= \frac{1}{2}\Tr \ee^{\slD_q^2/\Lambda^2} \ln (-\slD_q^2)\\
&= \frac{1}{2} \Tr f(-\slD_q^2)
\punc,
\end{align}
where \(f(x) = \ee^{-x/\Lambda^2}\ln x\). From \refeq{EqChiralGaugeIdentity}, we find
\beq
\frac{\dd}{\dd q}\slD_q = -\{ \slD_q, \beta \gamma^z \}\punc,
\eeq
and so, using the cyclicity of the trace to simplify the operator derivative,
\beq
\frac{\dd}{\dd q} \Tr f(-\slD_q^2) = -4 \Tr \left[ \beta \gamma^z g(-\slD_q^2) \right]
\eeq
where \(g(x) = x f'(x) = \ee^{-x/\Lambda^2}[1 - (x \ln x)/\Lambda^2]\).

As explained in \refsec{SecRegularizedContinuumDeterminant}, the trace \(\Tr\) is taken over functions with the correct boundary conditions, \refeq{EqpsipmPBCs}, and over matrix indices. For an operator \(K\), the trace can be defined by \(\Tr K = \int\dd^2\rv\,K(\rv,\rv)\), where \(K(\rv,\rv') = K \delta\super{\phiv}(\rv - \rv')\), with \(K\) acting on \(\rv\), is the kernel. Here, \(\delta\super{\phiv}\) is a periodic delta function with boundary phases \(\phiv\), analogous to \refeq{EqDefinePeriodicKroneckerDelta}; it is given by
\beq[EqPeriodicDelta]
\delta\super{\phiv}(\rv - \rv') = \frac{1}{{L_xL_y}} \sum_{\kv} \ee^{-\ii \kv \cdot (\rv-\rv')}\punc,
\eeq
where the sum includes the same set of \(\kv\) values as in \refeq{EqLambdaRegularizedDet}. We therefore have
\begin{widetext}
\beq
\frac{\dd}{\dd q} \Tr \ee^{\slD_q^2/\Lambda^2} \ln \slD_q =
-2\int \dd^2 \rv \, \beta(\rv) \frac{1}{{L_xL_y}} \sum_{\kv} \ee^{+\ii \kv \cdot \rv} \trace
\left[\gamma^z g(-\slD_q^2)\right] \ee^{-\ii \kv \cdot \rv}\punc{,}
\eeq
where \(\trace\) is the trace over just the matrix indices. Using \(\slD_q^2 = \Dv_q^2 1_2 + q F(\rv) \gamma^z\), where \(\Dv_q^2 = (\partial_\mu - \ii q a_\mu)(\partial_\mu - \ii q a_\mu)\) and \(F(\rv) = \epsilon_{\mu\nu} \partial_\mu a_\nu\), and
\beq
(\partial_\mu - \ii q a_\mu) \ee^{-\ii \kv\cdot \rv} = \ee^{-\ii \kv\cdot \rv}(\partial_\mu - \ii q a_\mu - \ii k_\mu)\punc,
\eeq
we find
\beq
\frac{\dd}{\dd q} \Tr \ee^{\slD_q^2/\Lambda^2} \ln \slD_q = -2\int \dd^2 \rv \, \beta(\rv) \frac{1}{{L_xL_y}} \sum_{\kv} \trace
\left[\gamma^z g\big({-(\Dv_q - \ii \kv)^2 1_2} - q F(\rv) \gamma^z\big)\right]\punc.
\eeq

For \(\Lambda \rightarrow \infty\) with \(x\) fixed, \(g(x) \rightarrow 1\), and so, because \(\trace \gamma^z = 0\), terms in the sum with \(\lvert \kv \rvert \ll \Lambda\) make no contribution. We can therefore replace the sum by an integral and expand \(g\) around \(-(-\ii \kv)^2 1_2 = \lvert \kv \rvert^2 1_2\), giving
\beq
\frac{\dd}{\dd q} \Tr \ee^{\slD_q^2/\Lambda^2} \ln \slD_q \approx -2\int \dd^2 \rv \, \beta(\rv) \int \frac{\dd^2 \kv}{(2\pi)^2} \trace
\left[
- q F(\rv) g'(\lvert\kv\rvert^2)1_2
\right]\punc,
\eeq
\end{widetext}
using \(\gamma^z\gamma^z = 1_2\) and omitting terms that vanish when taking the trace as well as those of higher order in \(\lvert \kv \rvert^{-1}\) (including from the commutator of \(\Dv_q\) and \(F\)). Making the substitution \(x = \lvert \kv \rvert^2\), this becomes
\beq
\frac{\dd}{\dd q} \Tr \ee^{\slD_q^2/\Lambda^2} \ln \slD_q \approx \frac{q}{\pi}\int \dd^2 \rv \, \beta(\rv)F(\rv) \int_0^\infty \dd x \, g'(x)
\eeq
for large \(\Lambda\). The integral over \(x\) is equal to \([g(x)]_0^\infty = -1\), and so \refeq{EqIntegralOverq} gives
\beq
\frac{I_{\phiv}[\av, 0]}{I_{\phiv}[\zerov,0]} = \exp \left[-\frac{1}{2\pi}\int \dd^2 \rv\, \beta(\rv)F(\rv)\right]
\punc.
\eeq
Using \(\epsilon_{\mu\nu}\epsilon_{\nu\rho} = -\delta_{\mu\rho}\), we can write \(F = \epsilon_{\mu\nu}\partial_\mu a_\nu = -\del^2 \beta\), and integrating by parts then gives \refeq{EqIphiaResult}.

\section{Chiral fermion bilinears and bosonic vertex operators}
\label{AppBosonization}

In this appendix, we show that \(\Delta_{\phiv}\), defined in \refeq{EqDefineDelta} as the expectation value of a product of chiral fermion bilinears \(\sigma_\pm\), can be expressed exactly in terms of the expectation value of a product of boson vertex operators, \(\ee^{\pm 2 \pi \ii h}\), as in \refeq{EqDeltaVertex}.

First, as noted after \refeq{EqDefineDelta}, \(\Delta_{\phiv}(\rv_1,\dotsc,\rv_n;\rv_1',\dotsc,\rv'_{n'}) = 0\) for \(n \neq n'\), because of the symmetry of the fermionic action under independent phase rotations of the two chiralities. Similarly, after decomposing \(h(\rv)\) as in \refeq{EqhDecomposition}, the only dependence on \(h_0\) is a factor \(\ee^{2\pi \ii (n-n') h_0}\), which gives \(\delta_{nn'}\) on integration. The fermionic and bosonic expressions are therefore both equal to zero for \(n\neq n'\), and so we restrict to \(n = n'\) in the rest of this appendix.

In \refsec{SecDerivation1}, we show that the fermionic and bosonic expressions are equal using an argument based on general properties of the correlators. (This method is the same in spirit as that of \refcite{Fujikawa2015}, but extended to allow for periodic boundary conditions. While we restrict to a rectangular system here, the arguments based on elliptic functions can straightforwardly be generalized to a parallelogram \cite{Lang1987}.) For completeness, we give explicit expressions for \(\Delta_{\phiv}\) in \refsec{SecBosonization2} based on the fermionic and bosonic theories, and show that they are equal.

\subsection{Derivation using correlator properties}
\label{SecDerivation1}

Restricting to \(n=n'\), we define
\begin{align}
\label{EqDeltaphif}
\Delta_{\phiv}\spr{f}(\rv_\cdot;\rv'_\cdot) &= \left\langle
\prod_{i=1}^n \sigma_-(\rv_i)\sigma_+(\rv_i')
\right\rangle_{\phiv}\\
\Delta_{\phiv}\spr{b}(\rv_\cdot;\rv'_\cdot) &=
\Lambda^{2n} 
\left\langle \prod_{i=1}^n \ee^{+2\pi \ii \h(\rv_i)} \ee^{-2\pi \ii\h(\rv_i')}\right\rangle_{\phiv}
\label{EqDeltaphib}
\punc,
\end{align}
where \((\rv_\cdot;\rv'_\cdot)\) stands for \((\rv_1,\dotsc,\rv_n;\rv_1',\dotsc,\rv_n')\). The expectation values are calculated in the continuum fermionic and bosonic distributions defined in \refeq{EqIphiah0} and after \refeq{EqDeltaVertex}, respectively (and reiterated below).

We will show that these functions both obey the following three conditions, and hence are equal. First,
\begin{widetext}
\beq[EqPoissonCondition]
\del_1^2 \ln \Delta_{\phiv}(\rv_\cdot;\rv'_\cdot) = -4\pi \sum_{j=1}^n \delta^2(\rv_1 - \rv'_j) + 4\pi \sum_{j=2}^n \delta^2(\rv_1 - \rv_j) + 4\pi\delta^2(\rv_1 - \Rv_{\phiv})\punc,
\eeq
where \(\del_1^2 = \left\lvert \frac{\partial}{\partial \rv_1} \right\rvert^2\) and
\beq[EqDefineRphi]
\Rv_{\phiv} \equiv \Rv_{\phiv}(\rv_2,\dotsc,\rv_n;\rv_1',\dotsc,\rv'_n) = \sum_{j=1}^n \rv'_j - \sum_{j=2}^n \rv_j + \frac{1}{2\pi}\left(-\phi_y L_x, \phi_x L_y\right)\punc,
\eeq
modulo boundary conditions. Second, for \(\lvert \rv_1 - \rv_j'\rvert \rightarrow 0\),
\beq[EqAsymptoticCondition]
\Delta_{\phiv}(\rv_1,\dotsc,\rv_n;\rv_1',\dotsc,\rv_n') \approx \frac{1}{(2\pi)^2\lvert \rv_1 - \rv_j'\rvert^2} \Delta_{\phiv}(\rv_2,\dotsc,\rv_n;\rv_1',\dotsc,\rv'_{j-1},\rv'_{j+1},\dotsc,\rv_n')
\punc,
\eeq
for \(j = 1,\dotsc,n\). On the right-hand side, \(\rv_1\) and \(\rv'_j\) do not appear in the arguments of \(\Delta_{\phiv}\); for \(n=1\) this function should be replaced by \(1\). (In the bosonic case, the delta functions are rounded over a length scale \(\Lambda^{-1}\) and the asymptotic expression is valid for \(\lvert \rv_1 - \rv_j'\rvert \gg \Lambda^{-1}\).) Finally, \(\Delta_{\phiv}\) obeys periodic boundary conditions under a shift of any argument by \(\Lv_\mu\). Given that \(\Delta\spr{f}_{\phiv}\) and \(\Delta\spr{b}_{\phiv}\) satisfy these conditions, \(\ln\frac{\Delta_{\phiv}\spr{f}(\rv_\cdot;\rv'_\cdot)}{\Delta_{\phiv}\spr{b}(\rv_\cdot;\rv'_\cdot)}\) is a harmonic function of \(\rv_1\) (i.e., a solution of Laplace's equation) with periodic boundary conditions and is therefore constant. The asymptotic form in \refeq{EqAsymptoticCondition} fixes the ratio to be \(1\) for all \(n\) by induction, and so the functions are equal.

\subsubsection{Fermionic theory}

Using \(\sigma_\pm(\rv) = \bar\psi_{\mp}(\rv)\psi_\pm(\rv)\), the fermionic correlator is
\beq
\Delta_{\phiv}\spr{f}(\rv_1,\dotsc,\rv_n;\rv'_1,\dotsc,\rv'_n) = (-1)^n G^+_{\phiv}(\rv'_1,\dotsc,\rv'_n;\rv_1,\dotsc,\rv_n)G^-_{\phiv}(\rv_1,\dotsc,\rv_n;\rv'_1,\dotsc,\rv'_n)
\eeq
where
\beq[EqFermion2npoint]
G^\pm_{\phiv}(\rv_1,\dotsc,\rv_n;\rv'_1,\dotsc,\rv'_n) = \left\langle \prod_{i=1}^n \psi_\pm(\rv_i) \bar\psi_\pm(\rv_i') \right\rangle_{\phiv}
\eeq
is the \(2n\)-point Green function for free fermions with action density \(\scL = \bar\psi_+ \partial_- \psi_+ + \bar\psi_- \partial_+ \psi_-\) and boundary phases \(\phiv\). From the definition in terms of Grassmann integrals, one can show that
\beq
G^-_{\phiv}(\rv_1,\dotsc,\rv_n;\rv'_1,\dotsc,\rv'_n)=(-1)^n\left[G^+_{\phiv}(\rv'_1,\dotsc,\rv'_n;\rv_1,\dotsc,\rv_n)\right]^*
\punc,
\eeq
and so
\beq
\Delta_{\phiv}\spr{f}(\rv_1,\dotsc,\rv_n;\rv'_1,\dotsc,\rv'_n) = \left\lvert G^+_{\phiv}(\rv'_1,\dotsc,\rv'_n;\rv_1,\dotsc,\rv_n) \right\rvert^2\punc.
\eeq
Using the boundary conditions on \(\psi_+\) and \(\bar\psi_{+}\), \refeq{EqpsipmPBCs}, it follows immediately that \(\Delta_{\phiv}\spr{f}\) has periodic boundary conditions, and so the third required condition is satisfied.

To show the other two conditions, we use the Dyson--Schwinger equation \cite{Zinn-Justin2002},
\beq[EqDysonSchwinger]
\partial_- G^+_{\phiv}(\rv'_1,\dotsc,\rv'_n;\rv,\rv_2,\dotsc,\rv_n) = \sum_{j=1}^n (-1)^{j}\delta^2(\rv - \rv_j') G^+_{\phiv}(\rv_1',\dotsc,\rv'_{j-1},\rv'_{j+1},\dotsc,\rv_n';\rv_2,\dotsc,\rv_n)\punc,
\eeq
\end{widetext}
where the function \(G^+_{\phiv}\) on the right-hand side should again by replaced by \(1\) for \(n=1\). (This is equivalent to using Wick's theorem and the Laplace expansion of the resulting determinant.) Using the distributional identity\footnote{\label{FootnoteWirtinger}Note that \(\partial_- = \partial_y - \ii \partial_x = -2\ii \partial_{z^*}\), where \(\partial_{z^*}\) is a Wirtinger derivative. For a justification of \refeq{EqDerivativeOfReciprocal}, see Section~5.2.1 of \refcite{DiFrancesco1997}.}
\beq[EqDerivativeOfReciprocal]
\partial_- \frac{1}{z} = -2\pi \ii \delta^2(z)
\punc,
\eeq
this implies that \(g(z) = G^+_{\phiv}(\rv'_1,\dotsc,\rv'_n;\rv,\rv_2,\dotsc,\rv_n)\), where \(z = x + \ii y\) for \(\rv = (x,y)\), is a meromorphic function with \(n\) first-order poles at \(z'_j\) (defined similarly in terms of \(\rv'_j\)), near which
\beq
g(z) \approx \frac{-(-1)^{j}}{2\pi\ii(z-z'_j)}
G^+_{\phiv}(\rv_1',\dotsc,\rv'_{j-1},\rv'_{j+1},\dotsc,\rv_n';\rv_2,\dotsc,\rv_n)\punc.
\eeq
It follows that \refeq{EqAsymptoticCondition} is satisfied.

To show \refeq{EqPoissonCondition}, consider also the zeros of \(g\). Due to fermionic antisymmetry, it has \(n-1\) first-order zeros at \(z_j\) for \(j = 2, \dotsc, n\). The boundary conditions on \(\bar\psi_+\) imply \(g(z + L_x) = \ee^{-\ii \phi_x}g(z)\) and \(g(z + \ii L_y) = \ee^{-\ii \phi_y}g(z)\), meaning that \(g\) is a quasi-elliptic (i.e., doubly quasi-periodic) meromorphic function. Since \(g'/g\) is then an elliptic and meromorphic, Liouville's theorems \cite{Lang1987} imply%
\footnote{Suppose \(g(z)\) has singularities of order \(m_\alpha\) at \(z = a_\alpha\), with \(m_\alpha > 0\) for zeros and \(m_\alpha < 0\) for poles. Integrating \(g'(z)/g(z)\) around the fundamental parallelogram, one finds that \(\sum_\alpha m_\alpha = 0\), as for an elliptic function \cite{DiFrancesco1997}. Doing the same with \(z g'(z)/g(z)\) gives
\[
2\pi \sum_\alpha m_\alpha a_\alpha = -L_x \phi_y + \ii L_y \phi_x\punc,
\]
up to shifts of \(\phi_\mu\) by \(2\pi\).}
that there is an \(n\)th first-order zero at \(z_{\phiv}\), corresponding to \(\Rv_{\phiv}\) in \refeq{EqDefineRphi}. Away from its poles, \(g(z)\) is an analytic function of \(z\), which implies that \(\ln \left[\lvert G^\pm_{\phiv}(\rv'_\cdot;\rv_\cdot)\rvert^2\right]\) is a harmonic function of \(\rv_1\), apart from at the points \(\rv_*\) corresponding to the poles and zeros of \(g\). Near these points,
\beq
\ln \left[\lvert G^\pm_{\phiv}(\rv'_\cdot;\rv_\cdot)\rvert^2\right]
\approx \pm \ln \left( \lvert \rv_1 - \rv_* \rvert^2 \right)\punc,
\eeq
with \(+\) for a zero and \(-\) for a pole. The right-hand side is proportional to the Green function for the Laplacian in 2D, and so \refeq{EqPoissonCondition} is satisfied.

\subsubsection{Bosonic theory}
\label{SecAppBosonicTheory}

We first rewrite the expectation value in \refeq{EqDeltaphib}, which is defined by \refeq{EqDefineVarphiEV}, using the decomposition, \refeq{EqhDecomposition}, into the zero mode \(h_0\), a real field \(\xi(\rv)\) with periodic boundary conditions and zero spatial average, and a pair of integers \(\Phiv\). As noted in the main text, \(\xi\) and \(\Phiv\) are independently distributed, and so
\beq[EqDeltaphib2]
\Delta_{\phiv}\spr{b}(\rv_\cdot;\rv'_\cdot) = \Omega_{\phiv}\left(\sum_{i=1}^n(\rv_i-\rv_i')\right)W_\Lambda(\rv_\cdot;\rv'_\cdot)\punc,
\eeq
where
\beq[EqDefineOmega]
\Omega_{\phiv}(\Rv) = \left\langle \ee^{-2\pi \ii\left( \frac{\Phi_y R_x}{L_x} - \frac{\Phi_x R_y}{L_y} \right)} \right \rangle_{\phiv}
\eeq
and
\beq[EqDefineDeltaxi]
W_\Lambda(\rv_\cdot;\rv'_\cdot) =
\Lambda^{2n} 
\left\langle \ee^{+2\pi \ii \sum_{i=1}^n [\xi(\rv_i) - \xi(\rv_i')]} \right\rangle
\punc.
\eeq
The variates \(\Phi_x\) and \(\Phi_y\) are integers with (complex) weight \(\omega_{\phiv}(\Phiv)\ee^{-\frac{\pi}{2}(\Phi_x^2/\rho + \Phi_y^2\rho)}\), where \(\omega_{\phiv}\) is defined in \refeq{EqDefineomegaphi}. The field \(\xi(\rv)\) has action density \(\frac{\pi}{2}(\del\xi)^2\), up to short-distance regularization such that \(W_\Lambda(\rv_\cdot;\rv'_\cdot)\) is independent of \(\Lambda\) for separations much larger than \(\Lambda^{-1}\) [see, e.g., \refeq{EqWepsilonAsymptotic}].

Both \refeqand{EqDefineOmega}{EqDefineDeltaxi} are clearly periodic under translation by \(\Lv_\mu\), and so \(\Delta\spr{b}_{\phiv}\) satisfies the third condition stated above.

The expectation value \(\Omega_{\phiv}(\Rv)\) can be expressed in terms of the function \(Q\) in \refeq{eq:resultForQ1} as \(\Omega_{\phiv}(\Rv) = Q(\phiv_{\Rv})/Q(\phiv)\), where \((\phiv_{\Rv})_\mu = \phi_\mu - 2\pi \sum_\nu \epsilon_{\mu\nu}R_\nu/L_\nu\). Using \refeq{EqQpde}, one finds
\begin{multline}
\left\lvert\frac{\partial}{\partial\Rv}\right\rvert^2 \ln Q(\phiv_{\Rv}) \\
= \frac{4\pi}{L_xL_y}\left[
\Sha\!\left(\frac{R_x}{L_x}+\frac{\phi_y}{2\pi}\right)\Sha\!\left(\frac{R_y}{L_y}-\frac{\phi_x}{2\pi}\right)
- 1\right]
\punc.
\end{multline}
For \(\Rv = \sum_i(\rv_i-\rv_i')\), the periodic delta functions select the point where \(\rv_1 = \Rv_{\phiv}\), and so
\beq[EqLaplacianLogOmega]
\del_1^2 \ln \Omega_{\phiv}\left(\sum_{i=1}^n(\rv_i-\rv_i')\right) = 4\pi \delta^2(\rv_1 - \Rv_{\phiv}) - \frac{4\pi}{L_x L_y}
\punc.
\eeq

The function \(W_\Lambda\) can be calculated using the result \(\left\langle \ee^{+2\pi\ii \Xi} \right\rangle = \ee^{-\frac{1}{2}(2\pi)^2\langle \Xi^2 \rangle}\) for a Gaussian variate \(\Xi\) with zero mean. Setting \(\Xi = \sum_{i=1}^n [\xi(\rv_i) - \xi(\rv_i')]\), we find, for \(n > 1\),
\beq[EqExpandxiVertexCorrelatorW]
W_\Lambda(\rv_\cdot;\rv'_\cdot) = \frac
{
\prod_{i,j=1}^n
W_\Lambda(\rv_i - \rv_j')
}
{
\prod_{1\le i < j \le n}
W_\Lambda(\rv_i - \rv_j)
W_\Lambda(\rv_i' - \rv_j')
}
\eeq
and, for \(n=1\),
\beq[EqGammaFromW]
W_\Lambda(\rv, \rv') = W_\Lambda(\rv-\rv')=\Lambda^2\ee^{-(2\pi)^2 \Gamma(\rv-\rv')}\punc,
\eeq
using translation invariance of \(\xi\).
The function \(\Gamma\) is defined by
\begin{align}\label{EqDefineGamma}
\Gamma(\rv-\rv') &= \frac{1}{2} \left\langle [\xi(\rv) - \xi(\rv')]^2 \right\rangle\\
&= \langle \xi(\zerov) \xi(\zerov)\rangle - \langle \xi(\rv)\xi(\rv')\rangle\punc,
\label{EqDefineGamma2}
\end{align}
which implies that it obeys the Poisson equation \cite{DiFrancesco1997}
\beq[EqGammaPoisson]
\pi \del^2 \Gamma(\rv) = \delta^2(\rv) - \frac{1}{L_xL_y}
\eeq
for \(\Lambda\rightarrow \infty\), with the constant due to the removed zero mode of \(\xi\). Rewriting \refeq{EqExpandxiVertexCorrelatorW} as
\begin{multline}
\label{EqExpandxiVertexCorrelator}
\frac{1}{(2\pi)^2}\ln\left\langle \ee^{+2\pi \ii \sum_{i=1}^n [\xi(\rv_i) - \xi(\rv_i')]} \right\rangle = -\sum_{i,j=1}^n \Gamma(\rv_i-\rv'_j)\\
+ \frac{1}{2}\sum_{\substack{i,j=1\\i \neq j}}^n \left[\Gamma(\rv_i-\rv_j) + \Gamma(\rv'_i-\rv'_j)\right]
\punc,
\end{multline}
acting with \(\del_1^2\), and combining the result with \refeq{EqLaplacianLogOmega}, we confirm that \(\Delta\spr{b}_{\phiv}\) satisfies \refeq{EqPoissonCondition}.

Finally, consider the asymptotic form of \(\Delta\spr{b}_{\phiv}\) as \(\rv_1 \rightarrow \rv'_j\). In the limit, \(\Omega_{\phiv}\!\left(\sum_i(\rv_i-\rv'_i)\right)\) is clearly given by removing \(\rv_1\) and \(\rv'_j\) from the set of arguments. Turning to \(W_\Lambda\), consider
\beq
\mathfrak{W}_{1j} = \frac{1}{(2\pi)^2}\ln \frac{\Lambda^{-2}W_\Lambda(\rv_1,\dotsc,\rv_n;\rv_1',\dotsc,\rv_n')}{W_\Lambda(\rv_2,\dotsc,\rv_n;\rv_1',\dotsc,\rv'_{j-1},\rv'_{j+1},\dotsc,\rv_n')}
\punc,
\eeq
which must have appropriate asymptotic behavior so that  \refeq{EqAsymptoticCondition} is satisfied. Using \refeq{EqExpandxiVertexCorrelator}, it is given by
\begin{multline}
\mathfrak{W}_{1j} = -\Gamma(\rv_1-\rv'_j) + \sum_{i=2}^n \left[\Gamma(\rv_1-\rv_i)-\Gamma(\rv_i-\rv'_j)\right] \\
+ \sum_{\substack{i=1\\i\neq j}}^n \left[\Gamma(\rv'_i-\rv'_j)-\Gamma(\rv_1-\rv'_i)\right]\punc.
\end{multline}
In both sums, every term is the difference of quantities with equal (and finite) limits as \(\rv_1 \rightarrow \rv'_j\), and so both vanish in this limit, leaving \(\mathfrak{W}_{1j} \approx -\Gamma(\rv_1-\rv'_j)\).

The asymptotic solution of \refeq{EqGammaPoisson} for \(\Lambda^{-1} \ll \lvert \rv \rvert \ll L_x, L_y\) is
\beq[EqGammaAsymptotic0]
\Gamma(\rv) \approx \frac{1}{(2\pi)^2} \ln \left[(2\pi)^2\lvert \rv \rvert^2\right] + \Gamma_0
\punc,
\eeq
where \(\Gamma_0\) is an arbitrary  constant that depends on the short-distance regularization. We fix the regularization (and hence \(\Gamma_0\)) by
\beq[EqGammaAsymptotic]
\Gamma(\rv) \approx \frac{2}{(2\pi)^2} \ln (2\pi\Lambda \lvert \rv \rvert)\punc,
\eeq
so that \refeq{EqAsymptoticCondition} is satisfied by \(W_\Lambda\), and hence by \(\Delta\spr{b}_{\phiv}\).

Note that, as stated after \refeq{EqDefineDeltaxi}, this choice of regularization ensures that \(W_\Lambda(\rv_\cdot;\rv'_\cdot)\) is independent of \(\Lambda\) for \(\lvert \rv_i - \rv_j'\rvert \gg \Lambda^{-1}\). For example, for \(n=1\),
\beq[EqWepsilonAsymptotic]
W_\Lambda(\rv) = \Lambda^2 \ee^{-(2\pi)^2\Gamma(\rv)} \approx \frac{1}{(2\pi)^2}\frac{1}{\lvert\rv\rvert^2}
\eeq
for \(\Lambda^{-1} \ll \lvert \rv\rvert \ll L_x,L_y\).

\subsection{Explicit expressions for correlation functions}
\label{SecBosonization2}

Wick's theorem gives the \(2n\)-point fermionic correlator, \refeq{EqDeltaphif}, as
\beq[EqDeltaWick]
\Delta\spr{f}_{\phiv}(\rv_\cdot;\rv'_\cdot) = (-1)^n \det\nolimits_{ij} G^+_{\phiv}(\rv_i'-\rv_j) \det\nolimits_{ij} G^-_{\phiv}(\rv_i-\rv_j')\punc.
\eeq
The two-point Green functions \(G^\pm_{\phiv}(\rv-\rv')\) depend only on the separation \(\rv - \rv'\) and are given by
\begin{align}
\label{EqFermionGreenFunction}
G^+_{\phiv}(\rv) &= \frac{\theta_1'(0,q)}{2L_y}\ee^{{\phi_y z}/{L_y}} \frac{\theta_1\!\left(\frac{1}{2}\varphi - \frac{\ii\pi}{L_y} z, q\right)}{\theta_1\!\left(\frac{1}{2}\varphi,q\right)\theta_1\!\left(-\frac{\ii\pi}{L_y} z, q\right)}\\
&\approx \frac{\ii}{2\pi  z}\qquad\text{for \(\lvert\rv\rvert \ll L_x,L_y\)}
\label{EqFermionGreenFunctionAsymptotic}
\end{align}
and \(G^-_{\phiv}(\rv) = -[G^+_{\phiv}(-\rv)]^*\), where \(z = x + \ii y\) for \(\rv = (x,y)\), \(\varphi = \phi_x + \ii \phi_y/\rho\), and \(q = \ee^{-\pi/\rho}\). We construct \(G^+_{\phiv}\) using a sum over Fourier modes in \refsec{AppFermionGreenFunction}, but one can easily confirm that it satisfies the required properties stated in \refapp{AppDecouplingGaugeField}. First, it is meromorphic in \(z\) and hence obeys \(\partial_- G^+_{\phiv}(\rv) = 0\) except at its singularities;\footnote{Note that we have defined \(\partial_{\pm} = \partial_y \pm \ii \partial_x\) (to match the Kasteleyn graph) but \(z = x + \ii y\), and so \(\partial_- z = 0\). This is also the reason for the unconventional factor of \(\ii\) in the asymptotic form, \refeq{EqFermionGreenFunctionAsymptotic}.} second, its asymptotic form, \refeq{EqFermionGreenFunctionAsymptotic}, near the singularity at \(\rv = \zerov\) obeys \(\partial_- G^+_{\phiv}(\rv) = \delta^2(\rv)\) near this point [see \refeq{EqDerivativeOfReciprocal}]; and finally, the quasiperiodicity of \(\theta_1\) \cite{NIST:DLMF} implies that it satisfies the required periodicity conditions.

Remarkably, the Frobenius--Stickelberger identity \cite[Eq.~(5.119)]{Krattenhaler2005}, which extends Cauchy's lemma \cite{Fujikawa2015} to periodic boundary conditions, allows the determinant to be written as
\begin{widetext}
\begin{multline}
\det\nolimits_{ij} G^+_{\phiv}(\rv'_i-\rv_j) =
\left[\frac{\theta_1'(0,q)}{2L_y}\right]^n
\ee^{-\frac{\phi_y}{L_y}\sum_i (z_i - z_i')}
\frac{\theta_1\!\left(\frac{1}{2}\varphi + \frac{\ii\pi}{L_y} \sum_i(z_i-z_i'), q\right)}{\theta_1\!\left(\frac{1}{2}\varphi,q\right)}\\\times
\frac{
\prod_{1\le i < j \le n} \theta_1\!\left(\frac{\ii\pi}{L_y}  (z_j-z_i), q\right)
\theta_1\!\left(\frac{\ii\pi}{L_y}  (z_i'-z_j'), q\right)
}
{
\prod_{i,j=1}^n
\theta_1\!\left(\frac{\ii\pi}{L_y}  (z_i-z_j'), q\right)
}
\punc.
\end{multline}
Substituting this into \refeq{EqDeltaWick}, we find
\beq[EqDelta2]
\Delta\spr{f}_{\phiv}(\rv_\cdot;\rv'_\cdot) =
\Omega_{\phiv}\left(\sum_i(\rv_i-\rv_i')\right)
\frac
{
\prod_{i,j=1}^n
W(\rv_i - \rv_j')
}
{
\prod_{1\le i < j \le n}
W(\rv_i - \rv_j)
W(\rv_i' - \rv_j')
}
\punc,
\eeq
\end{widetext}
(with the denominator defined to equal \(1\) in the case \(n=1\)) where
\beq[EqDefineOmega2]
\Omega_{\phiv}(\rv) = q^{2\left(\frac{x}{L_x}\right)^2}\ee^{-2x \phi_y/L_y}\left\lvert \frac{\theta_1\!\left(\frac{1}{2}\varphi + \frac{\ii \pi}{L_y}z, q\right)}{\theta_1\!\left(\frac{1}{2}\varphi, q\right)}\right\rvert^2\punc,
\eeq
and
\beq[EqDefineW]
W(\rv) = \frac{q^{-2\left(\frac{x}{L_x}\right)^2}}{(2L_y)^2}\left\lvert\frac{\theta_1'(0,q)}{\theta_1\!\left(\frac{\ii \pi}{L_y}z,q\right)}\right\rvert^2
\punc,
\eeq
which has the asymptotic form given in \refeq{EqWasymptotic} for \(\lvert\rv\rvert \ll L_x,L_y\).

One can show that \(\Omega_{\phiv}\) in \refeq{EqDefineOmega2} is equal to the function defined in \refeq{EqDefineOmega} by writing both in terms of \(Q\), as in \refsec{SecAppBosonicTheory}, and using
\refeq{eq:resultForQ2}. The function \(W(\rv)\) has periodic boundary conditions, obeys
\beq
-\del^2 \ln W(\rv) = 4\pi \delta^2(\rv) - \frac{4\pi}{L_x L_y}
\eeq
(see \refsec{SecBosonGreenFunctionDerivatives}), and has asymptotic form consistent with \refeq{EqWepsilonAsymptotic}; it is therefore equal to \(W_\Lambda\) defined in \refeq{EqGammaFromW} in the limit \(\Lambda\rightarrow \infty\). (We also construct the bosonic Green function explicitly in \refapp{AppBosonGreenFunction}, and hence confirm this statement.) It follows that the right-hand side of \refeq{EqDelta2} is equal to \(\Delta\spr{b}_{\phiv}\) in \refeq{EqDeltaphib2} in the same limit, which completes the proof that the fermionic and bosonic correlation functions are equal in this limit.

\section{Green functions}

\subsection{Fermions}
\label{AppFermionGreenFunction}

We want to calculate the Green function \(G^\pm_{\phiv}(\rv-\rv') = \langle\psi_{\pm}(\rv)\bar\psi_{\pm}(\rv')\rangle_{\phiv}\) for free fermions \(\psi_{\pm}\) with action density \(\bar\psi_{\pm}\partial_{\mp}\psi_{\pm}\) and quasiperiodic boundary conditions \(\psi_\pm(\rv + \Lv_\mu) = \ee^{\ii \phi_\mu} \psi_\pm(\rv)\) (and the complex conjugate for \(\bar\psi\)). This is the case \(n=1\) of the \(2n\)-point Green function defined in \refeq{EqFermion2npoint}, for which the Dyson--Schwinger equation, \refeq{EqDysonSchwinger}, reduces to \(\partial_{\mp} G^{\pm}_{\phiv}(\rv) = \delta\super{\phiv}(\rv)\), where \(\delta\super{\phiv}\) is a periodic delta function with boundary phases \(\phiv\), defined in \refeq{EqPeriodicDelta}. The Green function can therefore be written as
\beq[EqFermionGFsum]
G^\pm_{\phiv}(\rv) = \frac{1}{L_x L_y}\sum_{\kv} \frac{\ee^{\ii \kv \cdot \rv}}{\pm k_x + \ii k_y}\punc,
\eeq
where the sum is over the same wavevectors \(\kv\) as in \refeq{EqLambdaRegularizedDet}, i.e., \(k_\mu L_\mu = 2\pi n_\mu + \phi_\mu\) with \(n_\mu\) running over all integers. From this expression, we have \(G^-_{\phiv}(\rv) = -[G^+_{\phiv}(-\rv)]^*\), and so we restrict to \(G^+_{\phiv}\).

Applying the Poisson summation formula, \refeq{EqPSFfk}, to the sum over \(k_x\) gives
\beq
G^+_{\phiv}(\rv) = \frac{1}{L_y}\sum_{k_y} \ee^{\ii k_y y}
\sum_{m_x = -\infty}^\infty \ee^{-\ii m_x \phi_x} \int_{-\infty}^{\infty} \frac{\dd k_x}{2\pi} \frac{\ee^{\ii k_x (x + m_x L_x)}}{k_x + \ii k_y}
\punc.
\eeq
Using
\beq
\int_{-\infty}^{\infty} \frac{\dd k_x}{2\pi} \frac{\ee^{\ii k_x X}}{k_x + \ii k_y} = \begin{cases}
-\ii \ee^{k_y X} \sgn(k_y) & \text{for \(k_y X < 0\)}\\
0 & \text{for \(k_y X > 0\),}
\end{cases}
\eeq
the sum over \(k_y\) splits into two geometric series, which combine to give
\beq
G^+_{\phiv}(\rv) = -\frac{\ii}{L_y}
\ee^{\phi_y z/L_y}
\sum_{m_x = -\infty}^\infty \frac{\ee^{-\ii m_x \varphi}}{1 - \ee^{2\pi m_x / \rho}\ee^{2\pi z/L_y}}\punc,
\eeq
where \(z = x + \ii y\) for \(\rv = (x,y)\) and \(\varphi = \phi_x + \ii \phi_y / \rho\). (In spite of appearances, this expression has the required invariance under shifts of \(\phi_\mu\) by \(2\pi\).)

We finally use the Kronecker-type identity \cite{Mortenson2017}
\beq
\sum_{m = -\infty}^{\infty} \frac{\ee^{-2\ii m v}}{1 - q^{-2m}\ee^{2\ii w}}
= \frac{\ii}{2}
\frac{\theta_1'(0,q)\theta_1(v+w,q)}
{\theta_1(v,q)\theta_1(w,q)}\punc,
\eeq
with \(q = \ee^{-\pi/\rho}\), \(v = \frac{1}{2}\varphi\), and \(w = -\frac{\ii \pi}{L_y}z\), to give \refeq{EqFermionGreenFunction}.

\subsection{Bosons}
\label{AppBosonGreenFunction}

We want to calculate
\beq
\Gamma(\rv-\rv') = \frac{1}{2} \left\langle [\xi(\rv) - \xi(\rv')]^2 \right\rangle = \langle \xi(\zerov) \xi(\zerov)\rangle - \langle \xi(\rv)\xi(\rv')\rangle
\eeq
for a real scalar field \(\xi\) with action density \(\frac{\pi}{2}(\del\xi)^2\), periodic boundary conditions, and zero spatial average (i.e., with the \(\kv = \zerov\) mode removed). This function must be regularized using a short length scale \(\Lambda^{-1}\) and obeys the Poisson equation \refeq{EqGammaPoisson} in the limit \(\Lambda\rightarrow\infty\).

We do so by finding the Green function \(G\super{\xi}(\rv-\rv') = \langle \xi(\rv)\xi(\rv')\rangle\), which obeys the same Poisson equation but with a minus sign. In the same way as for the fermions, we express the Green function as a sum over wavevectors \(\kv\),
\beq
G\super{\xi}(\rv) = \frac{1}{L_x L_y}\sum_{\kv \neq \zerov} \frac{\ee^{\ii \kv \cdot \rv}}{\pi\lvert \kv \rvert^2}\punc,
\eeq
but now with \(k_\mu L_\mu \in 2\pi \dsZ\) and with the zero mode explicitly excluded from the sum.

Writing this as a sum over integers \(m_x\) and \(m_y\),
\beq
G\super{\xi}(\rv) = \frac{1}{(2\pi)^2}\frac{1}{\pi}\sum_{\mv \neq \zerov} \frac{\ee^{-2\pi\ii \left(\frac{m_xy}{L_y} - \frac{m_yx}{L_x}\right)}}{m_x^2/\rho + m_y^2\rho}
= -\frac{1}{(2\pi)^2}\ln Q(\Rv^\perp)
\punc,
\eeq
where \(Q\) is defined in \refeq{eq:defineQ} and \(R^\perp_\mu = 2\pi\sum_\nu\epsilon_{\mu\nu}r_\nu/L_\nu\). Using \refeq{eq:resultForQ2} for \(Q\) gives
\beq
G\super{\xi}(\rv) = \frac{1}{(2\pi)^{2}}\ln \left[q^{-2\left(\frac{x}{L_x}\right)^2}
\left\lvert\frac{\eta(q)}{\theta_1\!\left(\frac{\ii \pi}{L_y}z, q\right)}\right\rvert^2\right]\punc,
\eeq
and so
\beq[EqGxifromW]
G\super{\xi}(\rv) = \frac{1}{(2\pi)^{2}}\ln \frac{(2L_y)^2 \lvert\eta(q)\rvert^2 W(\rv)}{\lvert\theta_1'(0,q)\rvert^2}
\punc,
\eeq
where
\(W\) is defined in \refeq{EqDefineW}.

To find \(\Gamma\), we replace \(W\) by the regularized version
\beq
W_\Lambda(\rv) = \begin{cases}
W(\rv) & \text{for \(\lvert\rv\rvert \gg \Lambda^{-1}\)}\\
\Lambda^2 & \text{for \(\rv = \zerov\),}
\end{cases}
\eeq
and use \(\Gamma(\rv) = G\super{\xi}(\zerov) - G\super{\xi}(\rv)\) to give
\beq
\Gamma(\rv) = -\frac{1}{(2\pi)^{2}}\ln \frac{W_\Lambda(\rv)}{\Lambda^2}
\punc,
\eeq
which is equivalent to \refeq{EqGammaFromW}.

\subsection{Derivatives of bosonic field}
\label{SecBosonGreenFunctionDerivatives}

Correlations of \(\partial_\mu \xi\) can be expressed as
\beq
\langle \partial_\mu \xi(\rv) \partial_\nu \xi(\rv') \rangle = -\partial_\mu \partial_\nu G\super\xi(\rv - \rv')\punc.
\eeq
To find these, we start from \refeqand{EqGxifromW}{EqDefineW}, which give
\begin{multline}
(2\pi)^2 G\super\xi(\rv) = \frac{2\pi}{\rho}\frac{x^2}{L_x^2}
- \ln \theta_1\!\left(\frac{\ii\pi}{L_y}z,q\right)
- \ln \theta_1\!\left(\frac{\ii\pi}{L_y}z^*,q\right)\\
+ \text{const}
\punc,
\end{multline}
and take derivatives with respect to \(\partial_+ = 2\ii \partial_z\) and \(\partial_- = -2\ii \partial_{z^*}\). The results can be written in terms of the Weierstrass \(\wp\) function with periods \(2\omega_1 = \pi\) and \(2\omega_3 = \pi\ii/\rho\), which is given by \cite[Eq.~(23.6.14)]{NIST:DLMF}
\beq
\wp(w) = c(q) - \frac{\partial^2}{\partial w^2} \ln \theta_1(w,q)\punc,
\eeq
where
\beq
c(q) = \frac{\theta_1'''(0,q)}{3\theta_1'(0,q)}\punc.
\eeq
We find
\beq
\partial_+^2 G\super\xi(\rv) = \frac{1}{L_y^2} \left[ \wp\!\left(\frac{\ii \pi z}{L_y}\right) 
- c(q)
\right] - \frac{1}{\pi} \frac{1}{L_x L_y}
\eeq
and the same for \(\partial_-^2 G\super\xi\) but with \(z \rightarrow z^*\) (because \(\wp\) is meromorphic and even). Using \(\partial_+\partial_- = \del^2\) and \refeq{EqGammaPoisson} gives
\beq
\partial_+\partial_- G\super\xi(\rv) = -\frac{1}{\pi}\delta^2(\rv) + \frac{1}{\pi}\frac{1}{L_x L_y}
\punc.
\eeq

Writing \(\partial_x\) and \(\partial_y\) in terms of \(\partial_\pm\), we therefore find
\beq[EqdxiCorrelations]
\begin{aligned}
\langle \partial_x \xi(\rv) \partial_x \xi(\zerov) \rangle &= 
\frac{1}{2L_y^2} \re\wp\!\left(\frac{\ii \pi z}{L_y}\right) 
-\frac{c(q)}{2L_y^2}
-\frac{1}{\pi}\frac{1}{L_x L_y}\\
\langle \partial_y \xi(\rv) \partial_y \xi(\zerov) \rangle &= -\frac{1}{2L_y^2}\re\wp\!\left(\frac{\ii \pi z}{L_y}\right) 
+\frac{c(q)}{2L_y^2}\\
\langle \partial_x \xi(\rv) \partial_y \xi(\zerov) \rangle &= -\frac{1}{2L_y^2} \im\wp\!\left(\frac{\ii \pi z}{L_y}\right)
\punc,
\end{aligned}
\eeq
for \(\rv \neq \zerov\). In the each case, the constant term makes the integral around a period vanish, as required by the boundary conditions on \(\xi\). Using \(\wp(w) \approx w^{-2}\) for small \(w\), we find the asymptotic behavior given in \refeq{EqdxiCorrnsAsymptotic}.

\bibliography{dimersbibliography}

\end{document}